\documentstyle[12pt,epsfig]{article}

\newcommand{\stern}{+\hspace*{-0.327cm}\times}
\def\kreiso{\lower0.85pt\hbox{\Large $\bullet$}}
\def\kreisv{\raise0.85pt\hbox{$\scriptstyle\bigcirc$}}
\def\bbox{\lower0.85pt\hbox{$\Box$}}
\newcommand{\blacksquare}{\multiput(0,0)(0.25,0){22}{\line(0,1){6}}\;\,}
\def\lsim{\mathrel{\rlap{\lower4pt\hbox{\hskip1pt$\sim$}}
    \raise1pt\hbox{$<$}}}                
\def\gsim{\mathrel{\rlap{\lower4pt\hbox{\hskip1pt$\sim$}}
    \raise1pt\hbox{$>$}}}                

\begin{document}

\title{
\vspace{-2.5cm}
\flushleft{\normalsize DESY 97-125} \hfill\\
\vspace{-0.35cm}
{\normalsize HUB-EP-97/38} \hfill\\
\vspace{-0.35cm}
{\normalsize June 1997} \hfill\\
\vspace{0.5cm}
\centering{{\large \bf 
Scaling of Non-Perturbatively O(a) Improved Wilson
Fermions:}}\\
\hspace{-0.1cm} 
\centering{{\large \bf Hadron Spectrum, Quark Masses and Decay Constants}}
\\[1em]}
\author{\large M. G\"ockeler$^1$, R. Horsley$^2$, 
H. Perlt$^3$, P. Rakow$^4$,\\ G. Schierholz$^{4,5}$, A. Schiller$^3$
and P. Stephenson$^4$\\[2em]
        $^1$ Institut f\"ur Theoretische Physik, Universit\"at Regensburg,\\
                    D-93040 Regensburg,
                    Germany\\[0.5em]
        $^2$ Institut f\"ur Physik, Humboldt-Universit\"at,\\
                    D-10115 Berlin, Germany\\[0.5em]
        $^3$ Institut f\"ur Theoretische Physik, 
                    Universit\"at Leipzig,\\
                    D-04109 Leipzig, Germany\\[0.5em]
        $^4$ Deutsches Elektronen-Synchrotron DESY,\\ 
                    Institut f\"ur Hochenergiephysik und HLRZ,\\ 
                    D-15735 Zeuthen, Germany\\[0.5em]
        $^5$ Deutsches Elektronen-Synchrotron DESY,\\ 
                    D-22603 Hamburg, Germany}

\date{ }

\maketitle

\begin{abstract}
We compute the hadron mass spectrum, the quark masses and the meson 
decay constants in quenched lattice QCD with non-perturbatively $O(a)$ 
improved Wilson fermions. The calculations are done for two values of the
coupling constant, $\beta = 6.0$ and $6.2$, and the results are compared 
with the predictions of ordinary Wilson fermions. We find that the improved 
action reduces lattice artifacts as expected. 
\end{abstract}

\clearpage

\section{Introduction}

The calculation of hadron masses in lattice gauge theory has a long history.
Over the years there has been a steady improvement in computing power
and methods allowing simulations on larger volumes and smaller quark masses
with higher statistics. However, the progress to smaller lattice
spacing $a$ has been slower because of the high computer cost which
increases by at least a factor of $(1/a)^5$. Since it is so expensive
to reduce cut-off effects by reducing $a$, we should consider reducing
them by improving the action.

A systematic improvement program reducing the cut-off errors order by
order in $a$ has been proposed by Symanzik~\cite{Sym} and developed
for on-shell quantities in ref.~\cite{L&W}. The standard gluonic action has
discretization errors of $O(a^2)$, but those for Wilson fermions are
of $O(a)$. Therefore it is the fermionic action which is most in need of
improvement.
 
Sheikholeslami and Wohlert~\cite{S&W} proposed the action (we assume $r = 1$ 
throughout the paper)
\begin{equation}
S_F = S_F^{(0)} - \frac{\mbox{i}}{2} \, \kappa \, g \, c_{SW}(g) \, a \, a^4 
\sum_x \bar{\psi}(x) 
\sigma_{\mu\nu} F_{\mu\nu}(x) \psi(x),
\label{action}
\end{equation}
where $S_F^{(0)}$ is the original Wilson action and
\begin{equation}
F_{\mu\nu}(x) = \frac{1}{8 \mbox{i} g a^2} \sum_{\mu,\nu=\pm}
(U(x)_{\mu\nu}-U(x)^\dagger_{\mu\nu}). 
\label{F}
\end{equation}
In eq.~(\ref{F}) the sum extends over the four plaquettes in the 
$\mu\nu$-plane which have $x$ as one corner, and the plaquette 
operators $U(x)_{\mu\nu}$ are the products of the four link matrices 
comprising the plaquettes taken in a clockwise sense.
If $c_{SW}$ is appropriately chosen, this action removes all $O(a)$ errors
from on-shell quantities such as the hadron masses. A non-perturbative
evaluation of this function leads to~\cite{Letal2}
\begin{equation}
c_{SW}(g) = \frac{1 - 0.656 \, g^2 - 0.152 \, g^4 - 0.054 \, g^6}
{1 - 0.922 \, g^2}\:, \: g^2 \le 1 .
\label{nonpert}
\end{equation}
When we talk about improved fermions in the following, we always
understand that $c_{SW}$ has been chosen according to eq.~(\ref{nonpert}).

In this paper we shall present results for the light hadron mass spectrum,
the light and strange quark masses and the light meson decay constants
using the improved action. The calculation is done for two values of the
coupling, $\beta = 6.0$ and $6.2$, which allows us to test for scaling.

The mass calculations extend our earlier work~\cite{us1} where we have
examined the $c_{SW}$ dependence at $\beta = 6.0$. To exhibit the effect
of improvement, we have also done calculations with Wilson fermions on
the same lattices. Most of our Wilson data come from our structure function
calculations~\cite{Bi,Getal}, and we combine this with masses from 
the literature at
other $\beta$ values to see the dependence on $a$ clearly.

From the meson correlation functions we also extract meson decay constants
and quark masses. However, simply improving the action is not sufficient
to remove all $O(a)$ errors from these quantities. Here we also have to
improve the operators which is done by adding higher dimensional terms
with the same quantum numbers in an appropriate fashion.

This paper is organized as follows. In sec. 2 we briefly describe our
numerical method. The hadron masses are given in sec. 3, concentrating
in particular on the extrapolation to the chiral limit and the scaling
behavior of improved and Wilson action results. In sec. 4 we compute the
light and strange quark masses using two different methods, from the 
axial vector current Ward identity and from the lattice bare quark masses. The
meson decay constants are discussed in sec. 5. Finally, in sec. 6 we give our
conclusions.

\section{Computational Details}

Our calculations have mainly been done at $\beta = 6.0$ and $6.2$ on
$16^3 32$, $24^3 32$ and $24^3 48$ lattices. We use Quadrics (formerly
called {\it APE}) 
parallel computers. For the improved case the 
parameter $c_{SW}$ is given from eq.~(\ref{nonpert}) as $c_{SW} = 1.769$ at
$\beta = 6.0$ and $c_{SW} = 1.614$ at $\beta = 6.2$. The simulations are done
for at least five different $\kappa$ values in each case. This helps with
the extrapolation to the chiral limit.

For the gauge field update we use a combination of 16 overrelaxation sweeps
followed by a three-hit Metropolis update. This procedure is repeated 50
times to generate a new configuration. 

The improvement term in 
eq.~(\ref{action}) appears in the site-diagonal part of the action. The 
major overhead in our case is multiplication by this term during inversion
of the fermion mass matrix. In our basis of hermitean gamma matrices we
can rewrite this term as~\cite{peter}
\begin{eqnarray}
1 - \frac{\mbox{i}}{2} \kappa g c_{SW} \sigma \cdot F &=& 
                     \left( \begin{array}{cc}
                                            A & B \\
                                            B & A 
                                            \end{array}
                                     \right) \nonumber \\
&=& \frac{1}{2} \left( \begin{array}{rr}
                                            1 & -1 \\
                                            1 & 1 
                                            \end{array}
                     \right)
                     \left( \begin{array}{cc}
                                            A + B & 0 \\
                                            0 & A - B 
                                            \end{array}
                     \right)
                     \left( \begin{array}{rr}
                                            1 & 1 \\
                                            -1 & 1 
                                            \end{array}
                     \right)\, ,
\label{matr}
\end{eqnarray}
where $A$, $B$ are $6\times 6$ matrices (two-spinors with color), so that
instead of a $12\times 12$ multiplication we have two $6\times 6$
multiplications and two inexpensive coordinate transformations. This
reduces the overhead for the improvement in the inverter from 45\% to
30\%. Also, the inverse of the matrix in eq.~(\ref{matr}) is required on 
half the lattice due to the even-odd preconditioning. We now have to
invert two $6\times 6$ instead of a $12\times 12$ matrix. However, this is
only required once for each propagator inversion. 

For the matrix inversion
we mainly used the minimal residue algorithm, except for the lightest
quark mass on the larger lattices where we used the BiCGstab 
algorithm~\cite{BiCGstab,sesam}. As convergence criterion we chose
\begin{equation}
|r| \leq 10^{-6}
\end{equation}
for the residue, which is the best that can be achieved for our single
precision machine.

For the mass calculations we used Jacobi smearing for source and sink.
For a detailed description of our application of this procedure 
see ref.~\cite{Best}. 
We have two parameters we can use to set the size of our source, the number of
smearing steps, $N_s$, and the smearing hopping parameter, $\kappa_s$.
We chose $N_s = 50$ for $\beta = 6.0$ and $100$ for $\beta = 6.2$ and 
$\kappa_s = 0.21$ at both $\beta$ values. This gives roughly the same
r.m.s. radius in physical units in both cases, namely $0.4 \, \mbox{fm}$.
To define
the matrix elements for the decay constants and quark masses, we have also
computed correlation functions with smeared source and local sink. This
does not require any additional matrix inversions.

At $\beta = 6.0$ and $c_{SW} = 0$ we had generated $O(5000)$ configurations
for our structure function project on which we have computed the hadron
masses. To these we added $O(150)$ new configurations on which we computed
the meson decay constants and the chiral Ward identity.
For $c_{SW} = 1.769$ we have analyzed $O(1000)$ configurations. 
For the heavier quark
masses, $\kappa = 0.1487$ and
$\kappa = 0.1300$, $0.1310$, $0.1320$, respectively, the number of
configurations was $O(200)$. On the $24^3$ lattice we have 
generated $O(100)$ and $O(200)$ configurations at $c_{SW} = 0$ and $1.769$, 
respectively. 
At $\beta = 6.2$ we only ran on $24^3$ lattices. Here we have analyzed
$O(100)$ configurations for $c_{SW} = 0$ and $O(300)$ configurations for 
$c_{SW} = 1.614$.   
We employed both relativistic and non-relativistic wave 
functions~\cite{Bi,Getal}, except for the high statistics runs where we only
looked at the non-relativistic wave function in order to save computer time.

Besides our calculations at $\beta = 6.0$ and $6.2$ we also made
exploratory studies at $\beta = 5.7$ to see what effect varying $c_{SW}$
has on coarser lattices. If one decreases $\beta$, increases $c_{SW}$ or
increases $\kappa$, one starts to get problems with exceptional configurations.
This showed up in non-convergence of our fermion matrix inversions. It was,
however, only a real problem at $\beta = 5.7$, $c_{SW} = 2.25$ and~\cite{us1}
$\beta = 6.0$, $c_{SW} = 3.0$.

\section{Hadron Masses}

We consider hadrons where all the quarks have degenerate masses. We looked at
$\pi$, $\rho$, nucleon ($N$), $a_0$, $a_1$ and $b_1$ masses, and we 
have used this
nomenclature for all quark masses, not just in the chiral limit.

In our mass calculations we have made single exponential
fits to meson and baryon correlators over appropriate fit ranges.
The errors are determined using the bootstrap method with 50 data samples.
We present our hadron mass results in tables \ref{tm57}, \ref{tm60} and
\ref{tm62}. Table~\ref{tm60} updates the results presented in 
ref.~\cite{us1}. 
For the meson masses we found very little difference 
between using
relativistic and non-relativistic wave functions, and we settled for
relativistic wave functions (except for the high statistics runs). For the
nucleon we have chosen non-relativistic wave functions~\cite{Bi} 
which performed 
slightly better because the effective mass plateaus extended to larger times.
At $\beta = 6.0$ we repeated the lightest quark mass on $16^3 32$ on the
$24^3 32$ lattice, for both improved and Wilson fermions. The
values agree within less than 3\%. This indicates that all our results on the
$16^3 32$ lattice do not suffer from significant finite size effects.

\subsection*{{\it Chiral Behavior}}

To obtain the critical value of $\kappa$, $\kappa_c$, and the hadron masses
in the chiral limit, we extrapolate our data to zero $\pi$ mass. We first 
tried
\begin{equation}
m_{\pi}^2 = b \left(\frac{1}{\kappa} - \frac{1}{\kappa_c}\right).
\label{linear}
\end{equation}
Using this relation gives a rather poor fit of the data, and we 
saw
that there was a slight 
curvature in a plot of $m_{\pi}^2$ against $1/\kappa$.
Quenched chiral perturbation theory predicts~\cite{Sharpe}
\begin{equation}
m_{\pi}^2 = b' \left(\frac{1}{\kappa} - 
\frac{1}{\kappa_c}\right)^{\scriptstyle {\frac{1}{1+\delta}}},
\label{chiral}
\end{equation}
where $\delta$ is small and positive. We made fits using this formula but
found that $\delta$ was always negative. As in our previous work~\cite{us1} 
we conclude that our $\kappa$
values are too far from $\kappa_c$ for the formula to be applicable. This is
in agreement with observations made by other authors~\cite{Wein}. 
As an alternative parameterization of the curvature we used the 
phenomenological fit
\begin{equation}
\frac{1}{\kappa} = \frac{1}{\kappa_c} + b_2 m_{\pi}^2 + b_3 m_{\pi}^3.
\label{pheno}
\end{equation}
In table \ref{kappa} we give the values of $\kappa_c$ for the different 
fits. The
linear fits give $\chi^2/\mbox{dof}$ values of up to 40. The other two fits
both give acceptable values of $\chi^2$, but eq.~(\ref{pheno}) usually
gives a lower $\chi^2$ than eq.~(\ref{chiral}). In the following we shall 
take $\kappa_c$ from
the phenomenological fits.

\begin{figure}[h]
\vspace{1.0cm}
\begin{centering}
\epsfig{figure=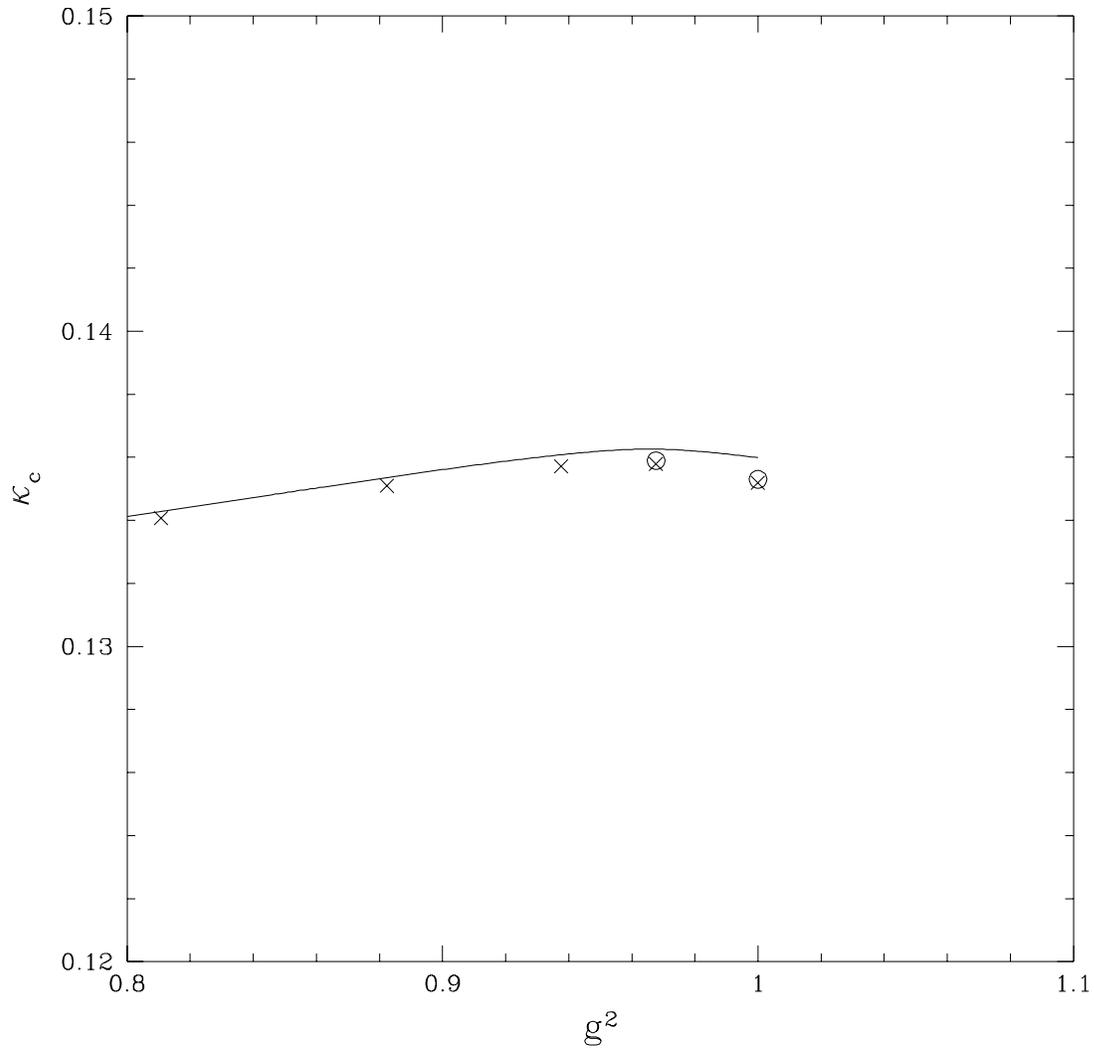,height=15.0cm,width=15.0cm}
\caption{The critical values of $\kappa$ for improved Wilson fermions as a 
function of $g^2$ from this work ($\kreisv$) and 
ref.~\protect\cite{Letal2} ($\times$). The curve is the tadpole improved result given in eq.~(\protect\ref{kappaimp}).}
\label{kappaplot}
\end{centering}
\vspace{-1.5cm} 
\end{figure}

In fig.~\ref{kappaplot} we plot $\kappa_c$ for improved Wilson 
fermions. We compare our results with the results of ref.~\cite{Letal2}. 
The agreement is excellent. 
In one-loop perturbation theory $\kappa_c$ is given by~\cite{us1} 
\begin{equation}
\kappa_c = \frac{1}{8}[1 + g^2 (0.108571 - 
0.028989 \, c_{SW} - 0.012064 \, c_{SW}^2)].
\end{equation}
The tadpole improved value of $\kappa_c$ that follows from this result is
\begin{equation}
\kappa_c = \frac{1}{8}[1 + g^{* \, 2} (0.025238 - 
0.028989 \, c_{SW} u_0^3 - 0.012064 \, (c_{SW} u_0^3)^2)] u_0^{-1},
\label{kappaimp}
\end{equation}
where $c_{SW}$ is given by eq.~(\ref{nonpert}),
\begin{equation}
u_0 = \langle \frac{1}{3} \mbox{Tr} U_\Box \rangle^{\frac{1}{4}}
\end{equation}
and $g^{* \, 2}$ is the boosted coupling constant defined by 
\begin{equation}
g^{* \, 2} = g^2/u_0^4.
\label{u0}
\end{equation}
In fig.~\ref{kappaplot} we compare the tadpole improved perturbative 
formula (\ref{kappaimp}) with the data where for the larger couplings we have 
taken $u_0$ from~\cite{L&M,Bali}. The curve and the data points agree within
less than 1\%. 
In eq.~(\ref{kappaimp}) one has the choice of using the lowest order tadpole
improved value of $c_{SW}$, namely $u_0^{-3}$~\cite{Letal2}, or the value
from eq.~(\ref{nonpert}) which is the value actually used in the simulations.
Both procedures remove all the tadpole diagrams and differ only by small
$O(g^4)$ terms, so they are both reasonable. We 
prefer the
second choice.

\begin{figure}
\begin{centering}
\epsfig{figure=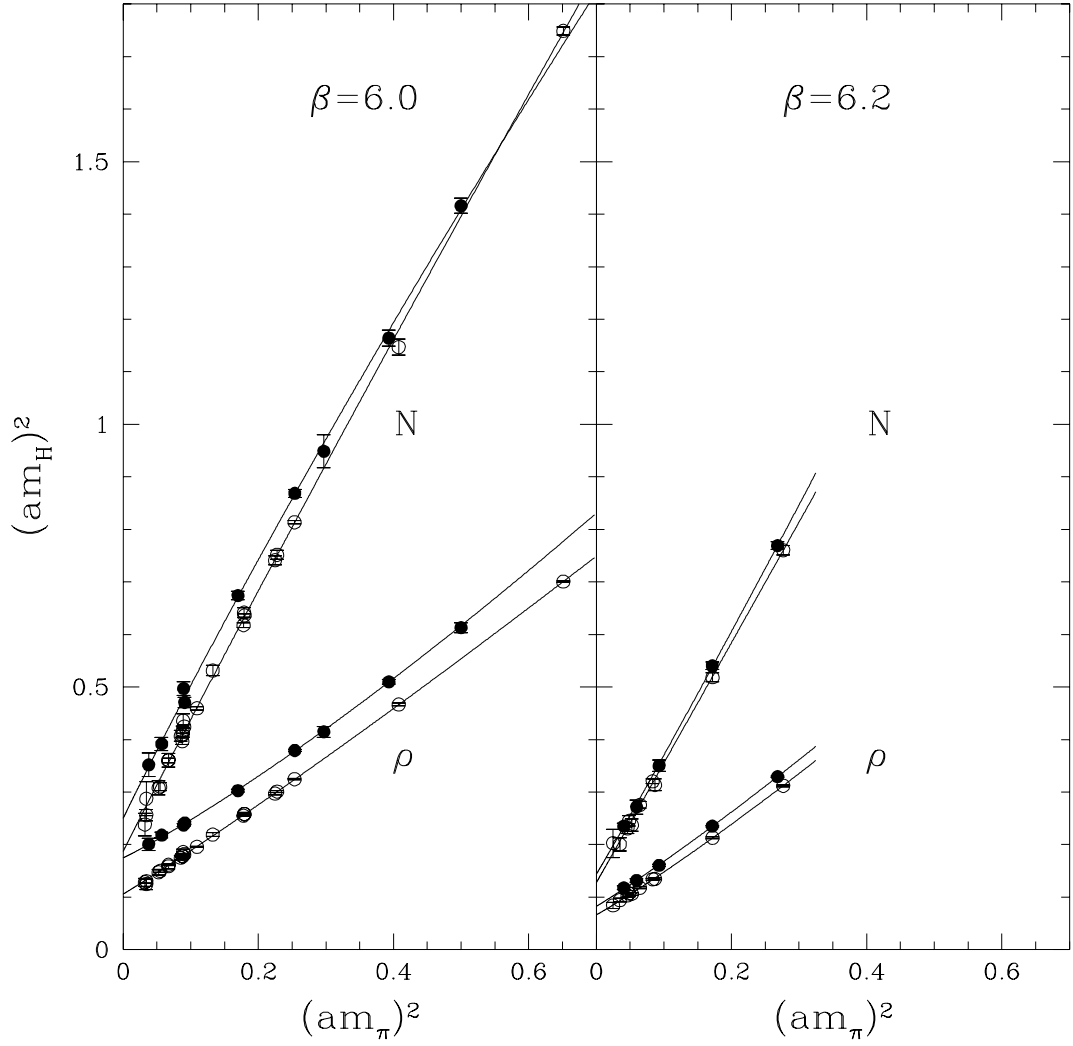,height=15.0cm,width=15.0cm}
\caption{Fits and chiral extrapolations of $\rho$ and nucleon masses
for improved ($\kreiso$) and Wilson
fermions ($\kreisv$).}
\label{mfit}
\end{centering}
\end{figure}

We fit the other hadron masses by the formula
\begin{equation}
m_H^2 = b_0 + b_2 m_{\pi}^2 + b_3 m_{\pi}^3 \, , \; 
H = \rho, N, \cdots.
\label{mesonfit}
\end{equation}
The result of the fit is shown in fig.~\ref{mfit} for both improved and 
Wilson fermion data. The Wilson fermion data are the world data compiled in
tables \ref{wilsonm1}, \ref{wilsonm2} and \ref{wilsonm3}.

We find this to be a more appropriate fit formula than the 
ansatz~\cite{Labrenz}
\begin{equation}
m_H = b'_0 + b'_2 m_{\pi}^2 + b'_3 m_{\pi}^3,
\end{equation}
because for the nucleon the plot of $m_N^2$ against $m_\pi^2$ (or $1/\kappa$)
is less curved than $m_N$ against $m_\pi^2$.
(Note that the two formulae differ only by terms of $O(m_{\pi}^4)$). 
To decide which fit formula is best and to do a reliable extrapolation to
the chiral limit, it is important to have many $\kappa$ values.
For the $a_0$, $a_1$ and $b_1$ masses only a 
two-parameter fit with $b_3$ set to zero was reasonable. The mass values in 
the chiral limit for our data are also given in tables \ref{tm57}, \ref{tm60} 
and \ref{tm62}.

We see that the effect of improvement is largest for the $\rho$ mass. In the 
chiral limit the difference between improved and Wilson results is 
25\% at $\beta = 6.0$ and still 12\% at $\beta = 6.2$. 
It is quite common to define the physical scale from the $\rho$ mass. The
relatively large change of this quantity from the Wilson to the improved
case suggests that it contains large $O(a)$ corrections, and that this 
procedure is misleading. A better procedure is to use the string tension or
$r_0$~\cite{r0}, the force parameter, as the scale.
For the nucleon mass the difference between the two actions is smaller.

\subsection*{{\it APE Plots}}

\begin{figure}
\begin{centering}
\epsfig{figure=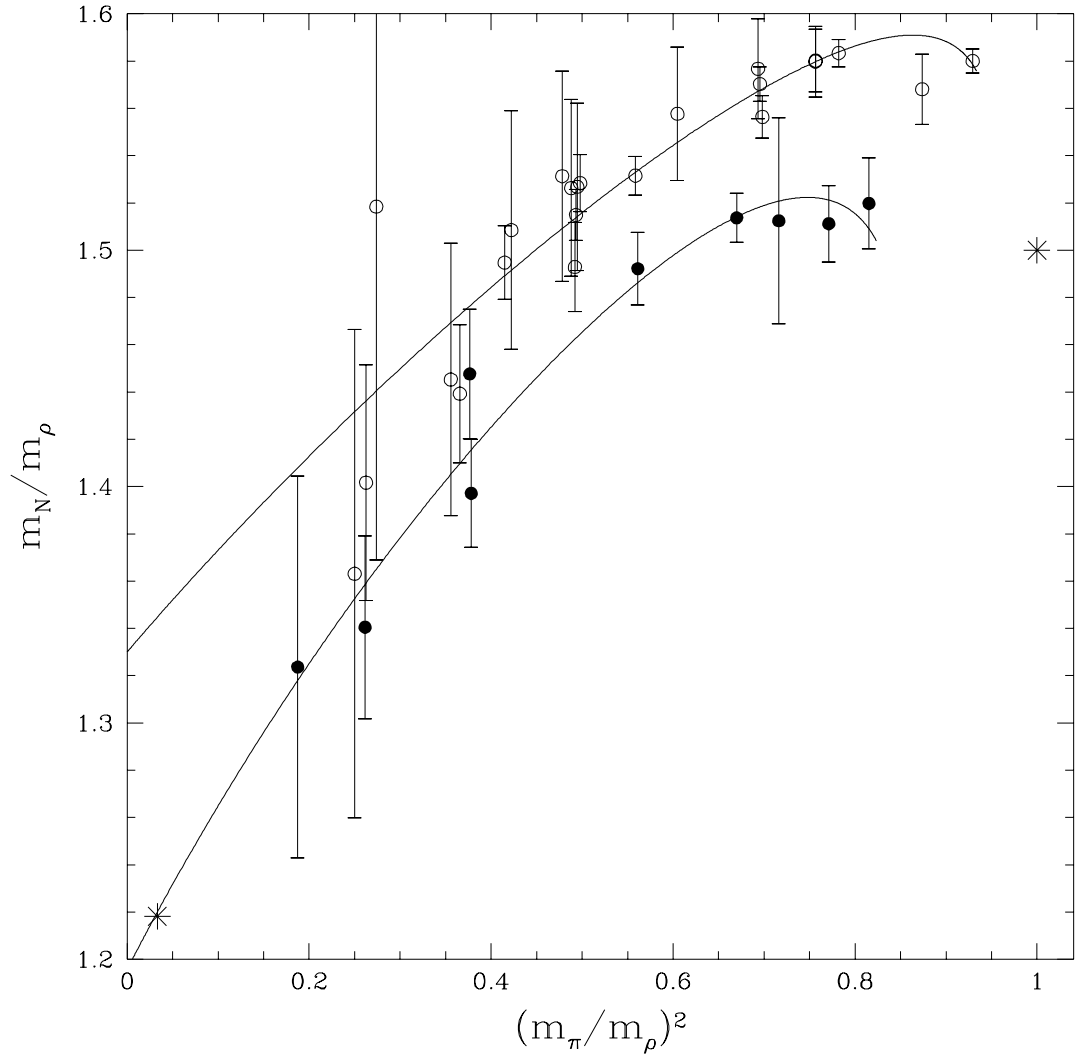,height=15.0cm,width=15.0cm}
\caption{$APE$ plot at $\beta = 6.0$ for improved ($\kreiso$) and Wilson
fermions ($\kreisv$) compared with the physical mass ratio ($\stern$) at the
physical quark mass and in the heavy quark limit. The solid lines are from
the mass fits described in the text.}
\label{ape6}
\end{centering}
\end{figure}

\begin{figure}
\begin{centering}
\epsfig{figure=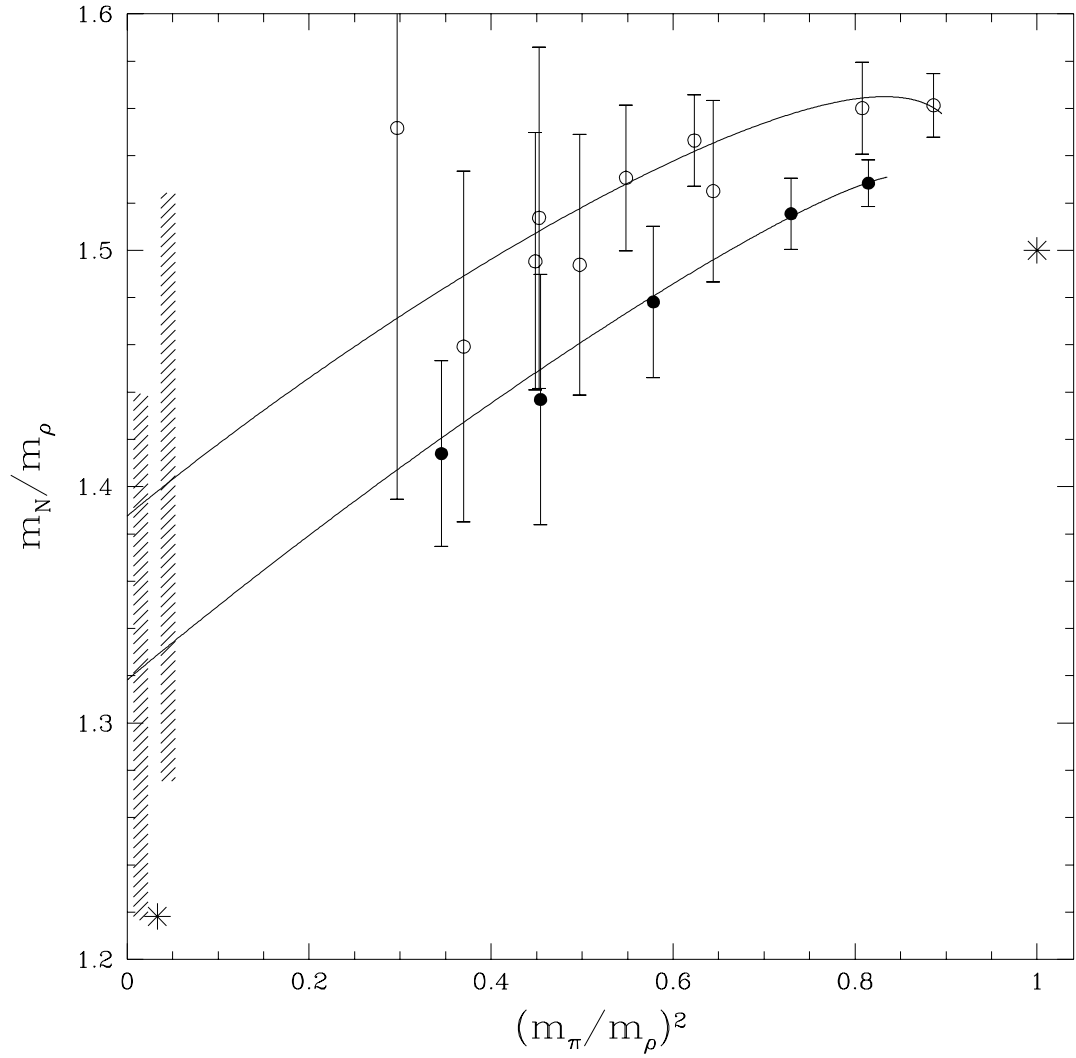,height=15.0cm,width=15.0cm}
\caption{The same as fig.~\ref{ape6}, but for $\beta = 6.2$. The hatched
bars indicate the errors in the chiral limit.}
\label{ape62}
\end{centering}
\end{figure}

In figs.~\ref{ape6} and \ref{ape62} we show the dimensionless ratio
$m_N/m_\rho$ as a function of $(m_\pi/m_\rho)^2$, a so-called {\it APE} plot, 
for $\beta = 6.0$ and
$6.2$, both for improved and Wilson fermions (the latter using the world data
given in tables \ref{wilsonm1} and \ref{wilsonm2}). 
The solid lines are the results of the ratio of the fits in fig.~\ref{mfit}.
At $\beta = 6.0$ we find that the mass ratio data are rather different
for the two actions. The improved results lie consistently lower than the
Wilson results. At $\beta = 6.2$ we find the same pattern in the data.

At $\beta = 6.0$ we can say something about the chiral limit. Our fits give
$m_N/m_\rho = 1.20(6)$ for improved fermions and 
$m_N/m_\rho = 1.33(2)$ for Wilson fermions. The improved results come closer
to the physical value than the Wilson results. At $\beta = 6.2$ we are 
lacking data at small quark masses and on larger volumes.
In the chiral limit our fits give $m_N/m_\rho = 1.32(11)$ for improved 
fermions and $m_N/m_\rho = 1.39(12)$ for Wilson fermions, so that we cannot 
say anything conclusive about the behavior of the two actions in the chiral 
limit in this case.

\subsection*{{\it Scaling Behavior}}

Let us now look at and compare the scaling behavior of the two actions. 
We shall limit our discussion to the $\rho$ mass because the errors of the 
nucleon are too large to make precise statements. In order to exhibit the 
cut-off effects most clearly, it has been suggested~\cite{Sommer} that 
$m_\rho$ should be plotted in units of the square-root of the string 
tension $K$ which has cut-off 
errors of $O(a^2)$ only. In table~\ref{stringdata} we have compiled the 
world string tension data. When there are several calculations, we performed
the weighted average.

\begin{figure}
\begin{centering}
\epsfig{figure=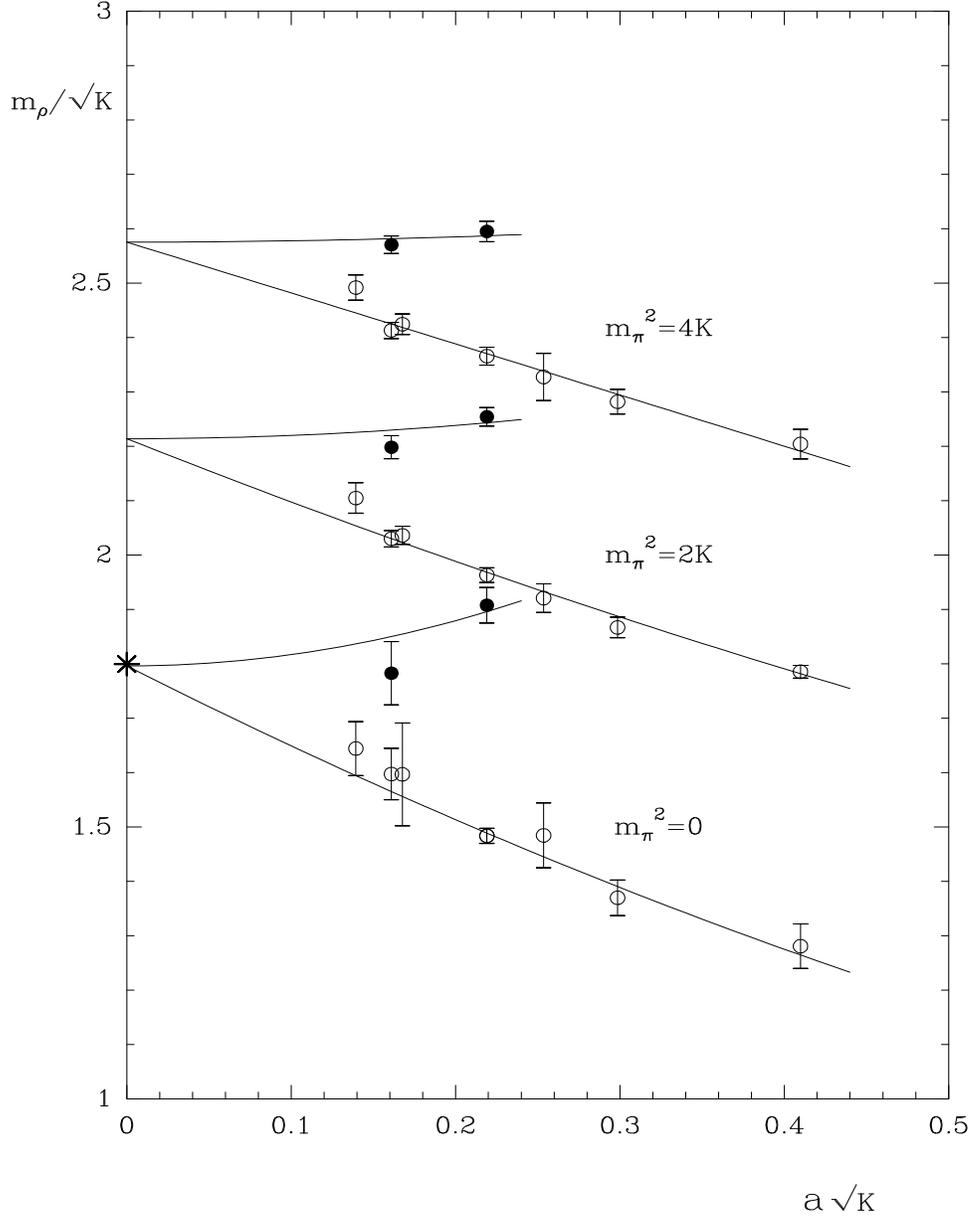,height=18.0cm,width=13.5cm}
\caption{The ratio $m_\rho/\protect\sqrt{K}$ as a function of the lattice 
spacing for improved ($\kreiso$) and Wilson fermions ($\kreisv$). The solid 
lines are from a simultaneous linear plus quadratic fit to the Wilson data 
and a quadratic fit to the improved data. This is compared with the 
experimental value ($\stern$) using $\protect\sqrt{K} = 427\, \mbox{MeV}$.}
\label{scalingplot}
\end{centering}
\end{figure}

In fig.~\ref{scalingplot} we plot the ratio
$m_\rho/\sqrt{K}$ as a function of $a \sqrt{K}$. This is done for fixed
physical $\pi$ masses with 
$m_\pi^2 = 0$, $2 K$ and $4K$. 
Comparing hadron masses at larger quark masses has 
the advantage
that this does not require large extrapolations of the lattice data but
rather involves small interpolations only. 
The Wilson fermion data shown are a fit to the
world data compiled in tables~\ref{wilsonm1}, \ref{wilsonm2} 
and \ref{wilsonm3}. 
As expected, the Wilson masses show 
practically a linear behavior in the lattice spacing $a$. We have done 
a simultaneous linear
plus quadratic fit to the Wilson data and a quadratic fit to
the improved data. 
The fit is constrained to agree in the continuum limit.
The result of the fit is shown by the solid lines in
fig.~\ref{scalingplot}. 
In the continuum limit we obtain $m_\rho/\sqrt{K} = 1.80(10)$. We compare 
this result with the experimental
$\rho$ mass. For the string tension we take the value
\begin{equation}
\sqrt{K} = 427\, \mbox{MeV}
\label{stringt}
\end{equation}
which has been obtained from a potential fit to the 
charmonium mass spectrum~\cite{Eetal}. Using this value the physical 
$m_\rho/\sqrt{K}$ is 1.80 which agrees with the lattice number.

As mentioned previously, an alternative scale from the potential is $r_0$. 
We have also compiled lattice
results for $r_0$ in table~\ref{stringdata}. We see that it scales very well
with $\sqrt{K}$, as the product $r_0 \sqrt{K}$ is approximately constant
at about $1.19$, while the lattice spacing $a$ changes by a factor of more
than five. However, the physical value of $r_0 \sqrt{K}$ is $1.06$, taking
$r_0^{-1}$ as $402\, \mbox{MeV}$ which follows from the same potential 
that gives
$\sqrt{K} = 427\, \mbox{MeV}$~\cite{Eetal}. It does not seem that this 
discrepancy will vanish as $a \rightarrow 0$. It is telling us that the lattice
potential has a slightly different shape to the continuum potential. This
may be an effect of quenching~\cite{campo}.

Although at $\beta = 5.7$ we do not know the correct value of $c_{SW}$,
using our larger value $c_{SW} = 2.25$ we find $m_\rho/\sqrt{K}$ = 1.94 in
the chiral limit. Comparing this number with fig.~\ref{scalingplot}, it 
indicates that $O(a^2)$ effects are moderate even at this coupling. 

\subsection*{{\it Mass Splitting}}

\begin{figure}
\begin{centering}
\epsfig{figure=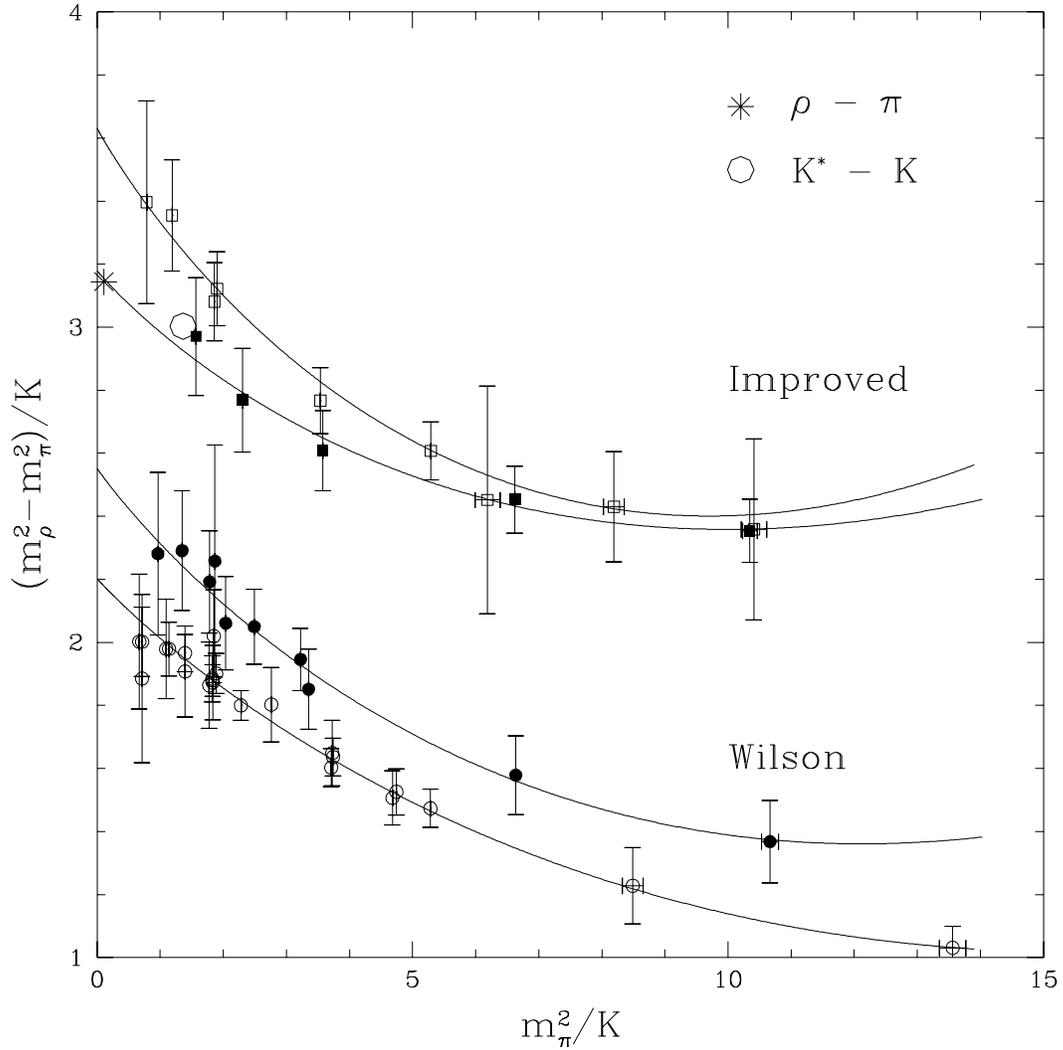,height=15.0cm,width=15.0cm}
\caption{The vector-pseudoscalar mass splitting as a function of the quark
mass. Open symbols correspond to $\beta = 6.0$, solid symbols to 
$\beta = 6.2$. This is compared with the physical 
$\rho$-$\pi$ ($\stern$) and $K^*$-$K$ (octogon) mass splitting.
The curves are from the mass fits.}
\label{rhopiplot}
\end{centering}
\end{figure}

The vector-pseudoscalar mass splitting
\begin{equation}
\Delta_{V-PS} = m_V^2 - m_{PS}^2
\end{equation}
is experimentally rather constant for all quark flavors. One finds
\begin{eqnarray}
m_\rho^2 - m_\pi^2 &=& 0.57\, \mbox{GeV}^2, \nonumber \\
m_{K^*}^2 - m_K^2  &=& 0.55\, \mbox{GeV}^2, \\ 
m_{D^*}^2 - m_D^2  &=& 0.55\, \mbox{GeV}^2. \nonumber 
\end{eqnarray}
Quenched lattice calculations with Wilson
fermions are unable to reproduce these numbers. 
Wilson fermions give a splitting which is much too small.
In fig.~\ref{rhopiplot} we
compare the experimental values of $m_\rho^2 - m_\pi^2$ and 
$m_{K^*}^2 - m_K^2$ with the lattice data and the mass fits. As before,
we have taken the string tension eq.~(\ref{stringt}) as the scale. 
In fig.~\ref{rhopiplot} we also show the results for improved
fermions and the corresponding mass fits as well. 
There is a noticeable change when going to
the improved case.
We find good agreement
with experiment for the absolute values.

In the heavy quark effective theory~\cite{Neubert}
\begin{equation}
\Delta_{V-PS} \propto \langle \bar{\Psi} \sigma_{\mu \nu} F_{\mu \nu} 
\Psi \rangle,
\end{equation} 
where $\Psi$ is the heavy quark field. So it is natural that turning on the
Sheikholeslami-Wohlert term would increase the mass splitting, and this is 
what we see.

\subsection*{{\it Wilson $\kappa_c$}}

\begin{figure}
\begin{centering}
\epsfig{figure=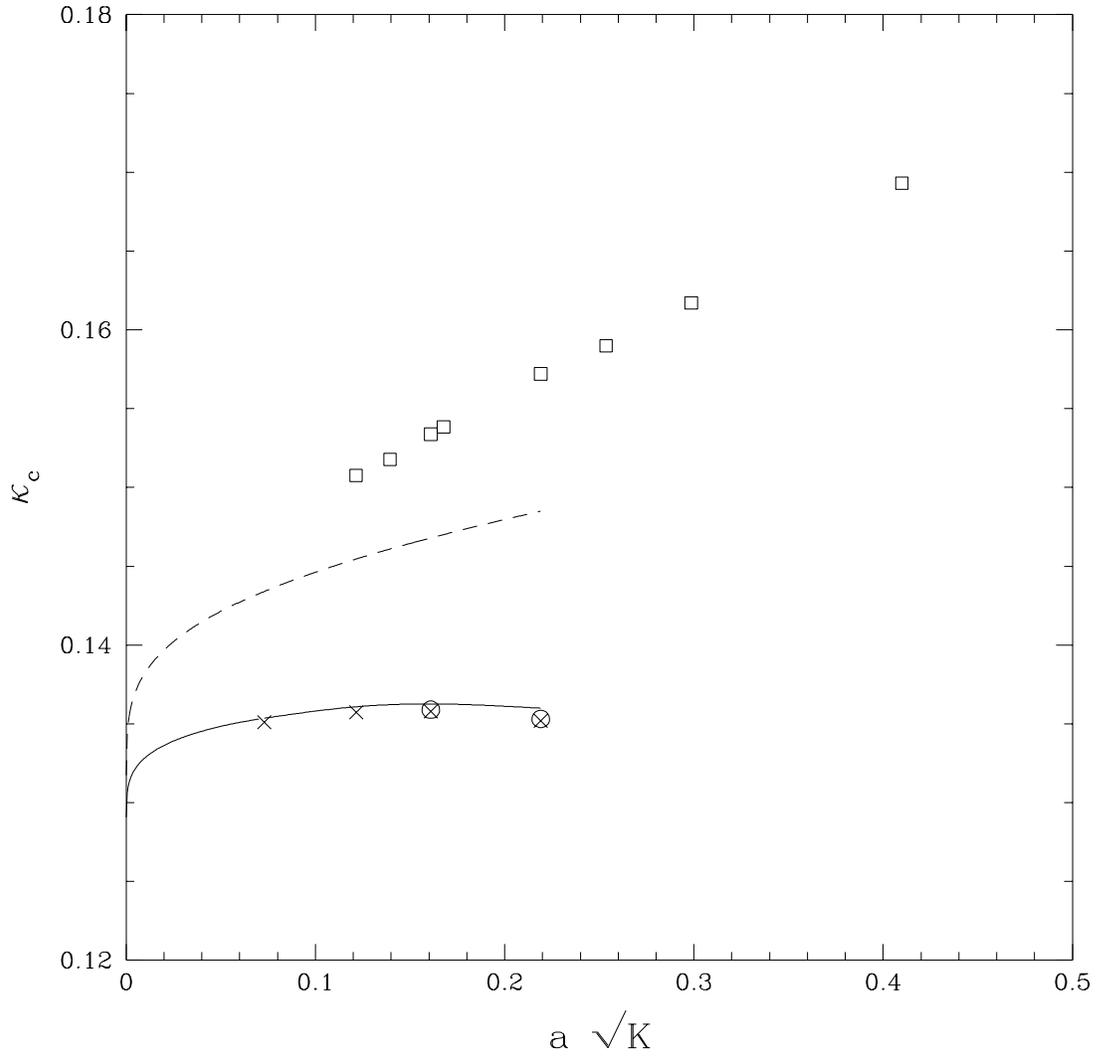,height=15.0cm,width=15.0cm}
\caption{The critical value of $\kappa$ as a function of the lattice 
spacing for Wilson fermions ($\bbox$).
The dashed curve is the prediction of tadpole improved perturbation theory.
This is compared with the results for improved fermions from 
fig.~\protect\ref{kappaplot}.}
\label{wkappa}
\end{centering}
\end{figure}

Let us now come back to the critical value of $\kappa$ for Wilson
fermions. In table~\ref{tkappa} we have given the values of $\kappa_c$
from a fit of the world data in tables~\ref{wilsonm1}, \ref{wilsonm2} and
\ref{wilsonm3} using the phenomenological ansatz (\ref{pheno}). In
fig.~\ref{wkappa} we plot these results as a function of $a \sqrt{K}$
(the string tension being taken from table~\ref{stringdata}). We see
that $\kappa_c$ is a linear function of $a$ over the whole range of the
data which extends from $\beta = 5.7$ to $6.4$. 
Comparing this with the improved $\kappa_c$, which is approximately constant,
we conclude that the Wilson $\kappa_c$ has large $O(a)$ effects.
We also compare the Wilson data with the predictions of tadpole improved
perturbation theory as given by eq.~(\ref{kappaimp}) with $c_{SW} = 0$.
Here we have taken the one-loop perturbative formula for $a$ beyond
$\beta = 6.8$ where there are no numerical values for the string tension  
available any more. Not even at the smallest value of $a$ can perturbation
theory describe the Wilson data. For improved fermions, on the other hand,
the agreement with tadpole improved perturbation theory is quite good, as we
have already noticed.

\section{Quark Masses}
\label{quark}

We shall now turn to the calculation of the quark masses. 
When chiral symmetry is dynamically broken, care has to be taken in
defining renormalized masses. In the continuum the
renormalized quark mass at scale $p^2 = \mu^2$ can be written~\cite{Pagels}
\begin{equation}
\frac{1}{4} \mbox{Tr} [S_F^{-1}(m) -S_F^{-1}(0)] = m(\mu),  
\end{equation}
where $S_F$ is the renormalized quark propagator which is to be evaluated in a
given gauge. This definition refers to the momentum subtraction scheme.
It is usual to give the quark masses in the $\overline{MS}$ scheme. To
convert from one scheme to the other, one has to go to high enough scales so
that one can use perturbation theory.
If the quark mass is defined in this way, then the renormalized mass is
proportional to the bare mass. 

On the lattice the standard assignment of the 
bare mass is
\begin{equation}
a m(a) = \frac{1}{2}\, (\frac{1}{\kappa} - \frac{1}{\kappa_c}),
\end{equation}
giving the renormalized mass as
\begin{equation}
m^{\overline{MS}}(\mu) = Z^{\overline{MS}}_m(a \mu, a m)\, m(a),
\end{equation}
where $Z^{\overline{MS}}_m(a \mu, a m)$ is the mass renormalization constant.
We call this method of determining the renormalized mass the standard method.

An alternative way of defining a bare mass is by means of the {\it PCAC} 
relation
between the divergence 
of the axial vector current $A_\mu = \bar{\psi}\gamma_\mu \gamma_5 \psi$ and
the pseudoscalar density $P = \bar{\psi}\gamma_5 \psi$,
\begin{equation}
\tilde{m}(a) = \frac{\partial_4 \langle A_4(x) {\cal O}\rangle}
{2 \langle P(x) {\cal O}\rangle},
\label{munimp}
\end{equation}
where ${\cal O}$ is a suitable operator having zero three-momentum and no
physical overlap with $A_4(x)$ and $P(x)$ to avoid contact terms. 
(See later on for 
a precise definition.) All operators are bare operators. To avoid anomaly
terms in eq.~(\ref{munimp}), flavor non-singlet operators are taken. We call
this method the Ward identity method. The renormalized mass is then given by
\begin{equation}
m^{\overline{MS}}(\mu) = 
\frac{Z_A(a m)}{Z^{\overline{MS}}_P(a \mu, a m)} \tilde{m}(a),
\end{equation}
where $Z_A(a m)$ and $Z^{\overline{MS}}_P(a \mu, a m)$ are the 
renormalization constants of the axial vector current and the pseudoscalar
density, respectively.

The quark mass inherits its scale dependence from the renormalization
constants $Z_m$ and $Z_P$ which involve logarithms of $\mu$. In the following
we will compute $Z_m$ and $Z_P$ perturbatively to one-loop order for lack
of a better, non-perturbative determination. To keep the
logarithms under control it is best to take $a \mu = 1$ and do the 
transformation to any other scale by the renormalization group formula
\begin{equation}
m^{\overline{MS}}(\mu') = \left( 
\frac{\alpha_s^{\overline{MS}}(\mu')}{\alpha_s^{\overline{MS}}(\mu)}
\right)^\frac{8}{22} m^{\overline{MS}}(\mu).
\label{rescale}
\end{equation}

In the continuum limit both procedures should give identical results
for $m^{\overline{MS}}(\mu)$. Note, however, that the two bare masses $m$ and
$\tilde{m}$ can be different, though they both vanish in the chiral limit.
On the lattice the two procedures may give different results for 
$m^{\overline{MS}}(\mu)$ due to non-universal discretization errors.
 
The lattice calculation of the quark masses now proceeds in two steps. In the 
first step one has to 
find the $\kappa$ values corresponding to the real world by adjusting (e.g.)
the pseudoscalar meson masses to their experimental numbers. In case of
the Ward identity method one furthermore has to compute $\tilde{m}$. 
In the second step the bare quark masses have to be converted to renormalized
masses. We shall compute the masses of the $u$ and $d$ quarks, which we assume
to be equal, and the mass of the strange ($s$) quark.   

\subsection*{{\it Improved Fermions}}

Let us consider the case of improved fermions first. Later on we shall 
compare our results with the predictions of Wilson fermions to
see the effect of improvement.

\begin{figure}
\vspace{ 4.5cm}
\begin{centering} \hspace{-3.25cm}
\epsfig{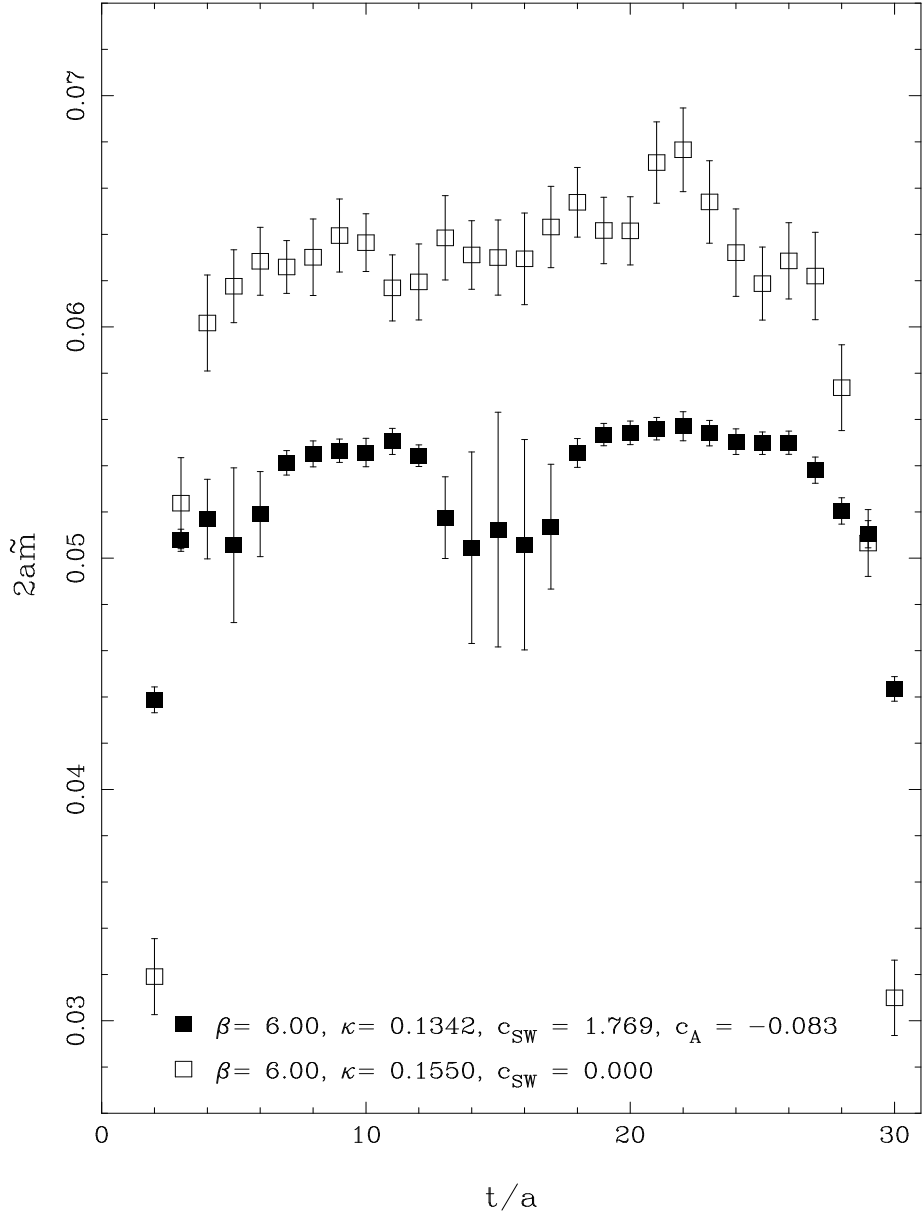}
\vspace{0.5cm}
\caption{The bare mass $\tilde{m}$ from the Ward identity method at 
$\beta = 6.0$ for Wilson ($\bbox$) and improved 
fermions ($\blacksquare$) on the $16^3 32$ lattice. The errors are bootstrap 
errors.}
\label{ward1}
\end{centering}
\end{figure}

\begin{figure}
\vspace{ 4.5cm}
\begin{centering} \hspace{-3.25cm}
\epsfig{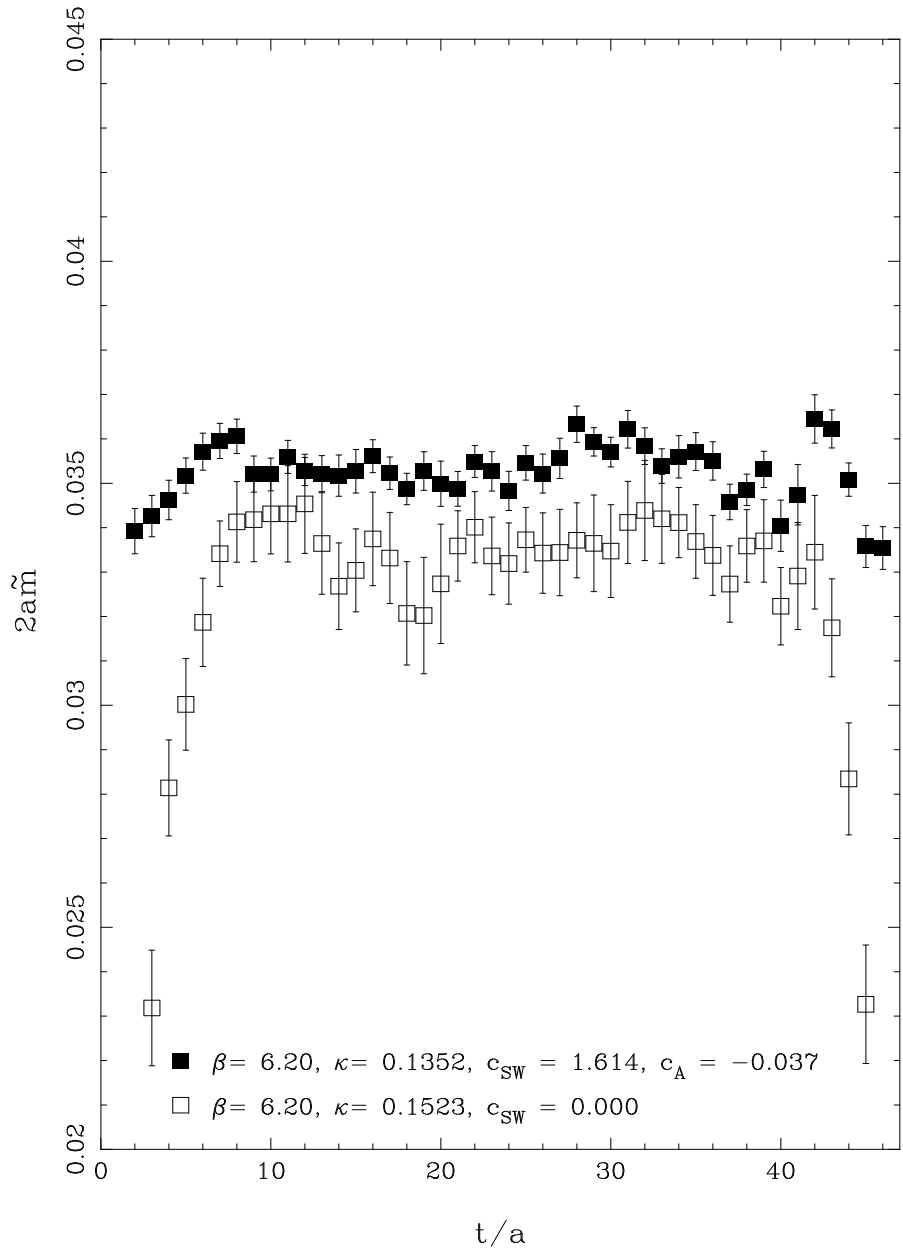}
\vspace{0.5cm}
\caption{The same as fig.~\protect\ref{ward1} but for 
$\beta = 6.2$ for Wilson ($\bbox$) and improved 
fermions ($\blacksquare$).}
\label{ward2}
\end{centering}
\end{figure}

We will discuss the Ward identity method first. For the operator ${\cal O}$ 
we take the pseudoscalar density
\begin{equation}
P(0) = \sum_{\vec{x}} P(x_4=0,\vec{x})
\end{equation}
and smear it as we did in the hadron mass calculations.
As the $P(0)$ part is common to all two-point functions, we could have
used any operator projecting onto the pseudoscalar state.
Similarly, we write
\begin{equation}
A_4(t) = \sum_{\vec{x}} A_4(x_4=t,\vec{x}).
\end{equation}
For improved fermions the axial vector current in 
eq.~(\ref{munimp}) is to be replaced by
\begin{equation}
A_4 \rightarrow A_4 + c_A a \partial_4 P(x),
\end{equation}
where $c_A$ is a function of the coupling only. 
The time derivative $\partial_4$ is taken to be the average of the forward and 
backward derivative.  
The coefficient $c_A$ has 
been computed in~\cite{Letal2} giving $c_A = -0.083$ at $\beta = 6.0$
and $c_A = -0.037$ at $\beta = 6.2$. The resulting bare mass
\begin{equation}
\tilde{m}(a) = \frac{\partial_4 \langle A_4(t) P(0)\rangle 
+ c_A a \partial_4^2 \langle P(t) P(0)\rangle}
{2 \langle P(t) P(0)\rangle}
\label{mward1}
\end{equation}
has been plotted in fig.~\ref{ward1} for $\beta = 6.0$ and our smallest
quark mass on the $16^3 32$ lattice.
In fig.~\ref{ward2} we show the same quantity for $\beta = 6.2$ and 
our smallest quark mass on the $24^3 48$ lattice.
(Also shown in these figures are the results for Wilson fermions which
we will discuss later on.) 
Equation~(\ref{mward1}) should be independent of $t$, except where the
operators physically overlap with the source, if the cut-off effects have been
successfully removed. In both cases, but in particular at $\beta = 6.2$,
we see a smaller deviation from the plateau at small and large $t$ values.
To obtain the mass, we fit the ratio (\ref{mward1}) to a constant. We
have used the same fit ranges as for the pion mass. 
The results of the fit are given in 
tables~\ref{tward1},~\ref{tward2}. At $\beta = 6.0$ in the improved case we 
see that at $\kappa = 0.1342$ we have small finite size effects, indicating
again that our results on the $16^3 32$ lattice are not significantly
volume dependent.

For both the Ward identity and the standard method we choose to determine the
$\kappa$ values from the pseudoscalar meson masses.
Sometimes the $\phi(1020)$ meson is taken for the determination of the
strange quark mass. However, we do not think that this is a good idea because
of potential $\omega-\phi$ mixing~\cite{particle_table}. We generalize 
eq.~(\ref{pheno}) to the case of two different quark masses by writing
\begin{equation}
\frac{1}{2} \, (\frac{1}{\kappa_1} + \frac{1}{\kappa_2}) - \frac{1}{\kappa_c}
= b_2 m_{PS}^2 + b_3 m_{PS}^3
\label{twokappa}
\end{equation}
with the same coefficients $b_2$, $b_3$ as before. 
This is inspired by chiral perturbation theory where it is expected that the 
pseudoscalar mass is a function of the sum of quark and antiquark mass,
$m_q + m_{\bar{q}}$, even when quark and antiquark have different flavors.
By fixing $m_{PS}$ to 
the physical pion mass $m_{\pi^\pm}$, using the string tension values compiled
in table~\ref{stringdata} with eq.~(\ref{stringt}) as the scale, we 
find the value for $\kappa_{u,d} = \kappa_1 \equiv \kappa_2$. The strange 
quark mass is
obtained by identifying $m_{PS}$ with the kaon mass $m_{K^\pm}$, taking 
$\kappa_1 = \kappa_{u,d}$ as input and solving for $\kappa_2 = \kappa_s$. 
This gives for the light mass
\begin{equation}
m_{u,d} a = \frac{1}{2} \, \left(\frac{1}{\kappa_{u,d}} - 
\frac{1}{\kappa_c}\right) = \left\{ \begin{array}{ll}
                                    0.001836(36) & \mbox{for}\; \beta = 6.0, \\
                                    0.001384(36) & \mbox{for}\; \beta = 6.2.
                                    \end{array}
                            \right.
\label{mmud}
\end{equation}
For the strange mass we get
\begin{equation}
m_s a = \frac{1}{2} \, \left(\frac{1}{\kappa_s} - 
\frac{1}{\kappa_c}\right) = \left\{ \begin{array}{ll}
                                    0.0419(11) & \mbox{for}\; \beta = 6.0, \\
                                    0.0310(11) & \mbox{for}\; \beta = 6.2,
                                    \end{array}
                            \right.
\label{mms}
\end{equation}
where $m_{u,d} = 1/2 \, (m_u + m_d)$.

The bare masses $\tilde{m}_{u,d}$, $\tilde{m}_s$ are computed analogously.
We write
\begin{equation}
\tilde{m} \equiv \frac{1}{2} (\tilde{m}_1 + \tilde{m}_2) = 
\tilde{b}_2 m_{PS}^2 + \tilde{b}_3 m_{PS}^3.
\end{equation}
Using this parameterization we first fit the masses in 
tables~\ref{tward1},~\ref{tward2} to the pseudoscalar masses in 
tables~\ref{tm60},~\ref{tm62}. This gives us $\tilde{b}_2$, $\tilde{b}_3$.
We then determine $\tilde{m}_{u,d}$, $\tilde{m}_s$ by fixing $m_{PS}$ to
the physical pion and kaon masses, respectively, as before.

The mass dependence of the renormalization constant $Z_A(am)$ can be 
parameterized as~\cite{Letal3}
\begin{equation}
Z_A(am) = (1 + b_A am) Z_A. 
\end{equation}
The renormalization constant $Z_A$ has been computed non-perturbatively in
ref.~\cite{Letal4}. The fit formula in this paper gives
$Z_A = 0.7924$ at $\beta = 6.0$ and
$Z_A = 0.8089$ at $\beta = 6.2$. The coefficient $b_A$ is only known 
perturbatively to one-loop order~\cite{Sint}. The best we can do at present 
is to take the tadpole improved value. For the boosted coupling we use
$\alpha_s^{\overline{MS}}(1/a)$, giving
\begin{equation}
b_A = 1 + \alpha_s^{\overline{MS}}(1/a)\, 1.912,
\end{equation}
where we take $\alpha_s^{\overline{MS}}(1/a) = 0.1981$ at $\beta = 6.0$ and 
$\alpha_s^{\overline{MS}}(1/a) = 0.1774$ at $\beta = 6.2$~\cite{L&M}. For 
$Z_P^{\overline{MS}}(a \mu, am)$ we write
\begin{equation}
Z_P^{\overline{MS}}(a \mu, am) = (1 + b_P am) Z_P^{\overline{MS}}(a \mu).
\end{equation}
The renormalization constant $Z_P^{\overline{MS}}(a \mu)$ has been computed 
perturbatively~\cite{C&us}. The result is
\begin{equation}
Z_P^{\overline{MS}}(a \mu) = 1 - \frac{g^2}{16 \pi^2} C_F (-6 \ln(a \mu)
+22.595 -2.249 c_{SW} + 2.036 c_{SW}^2),
\label{zp}
\end{equation}
with $C_F = 4/3$. We shall take the scale $\mu = 1/a$ and use the 
tadpole improved value of eq.~(\ref{zp})
which turns out to be
\begin{equation}
Z_P^{\overline{MS}}(a \mu = 1) = \left[ 1 - 
\frac{\alpha_s^{\overline{MS}}(1/a)}{4 \pi}
\left(16.967 -2.999 c_{SW} u_0^3 + 2.715 (c_{SW} u_0^3)^2\right) \right] u_0.
\label{zpimp}
\end{equation}
(We use $u_0 = 0.8778$ at $\beta = 6.0$ and 
$u_0 = 0.8851$ at $\beta = 6.2$). The coefficient $b_P$ has also been
computed perturbatively to one-loop order~\cite{Sint}.
Again we shall use the tadpole improved value
\begin{equation}
b_P = 1 + \alpha_s^{\overline{MS}}(1/a)\, 1.924.
\end{equation}

We have also computed the renormalization constants $Z_A(am)$, 
$Z_P(a \mu, am)$
non-per\-tur\-ba\-tively~\cite{Oetal2}. So far we have results for 
$\beta = 6.0$ only.
Our numbers are in fair agreement with the non-perturbative calculation
in ref.~\cite{Letal4} and the tadpole improved value (\ref{zpimp}).
However, for small $\mu$ the constant $Z_P$ behaves very differently from
the perturbative formula.

To compare the results at the two different $\beta$ values, we rescale
them both to $\mu' = 2\, \mbox{GeV}$ using formula (\ref{rescale}).
As before, we use the string tension to convert the lattice spacing
into physical units. The resulting quark masses 
$m_{u,d}^{\overline{MS}}(2\, \mbox{GeV})$, 
$m_s^{\overline{MS}}(2\, \mbox{GeV})$
are given in table~\ref{mMeV}.

Let us now discuss the standard method. We already have determined
$m_{u,d}(a)$, $m_s(a)$ in eqs.~(\ref{mmud}), (\ref{mms}). For the 
renormalization constant 
$Z_m^{\overline{MS}}(a \mu, am)$ we write
\begin{equation}
Z_m^{\overline{MS}}(a \mu, am) = (1 + b_m am) Z_m^{\overline{MS}}(a \mu).
\end{equation}
The constant $Z_m^{\overline{MS}}(a \mu)$ has been computed
perturbatively~\cite{C&us}. We obtain
\begin{equation}
Z_m^{\overline{MS}}(a \mu) = 1 - \frac{g^2}{16 \pi^2} C_F (6 \ln(a \mu)
- 12.952 - 7.738 c_{SW} + 1.380 c_{SW}^2).
\end{equation}
The tadpole improved value, which we will be using, is
\begin{equation}
Z_m^{\overline{MS}}(a \mu = 1) = \left[ 1 - 
\frac{\alpha_s^{\overline{MS}}(1/a)}{4 \pi}
\left(- 4.110 - 10.317 c_{SW} u_0^3 + 1.840 (c_{SW} u_0^3)^2 \right) \right] 
u_0^{-1}.
\label{zmimp}
\end{equation}
The coefficient $b_m$ has been computed in~\cite{Sint}. The tadpole improved
value is
\begin{equation}
b_m = -\frac{1}{2} - \alpha_s^{\overline{MS}}(1/a) \, 1.210.
\end{equation}
Again we extrapolate the quark masses to $\mu' = 2\, \mbox{GeV}$ using
eq.~(\ref{rescale}). The results which follow from this approach
are listed in 
table~\ref{mMeV} as well.
 
\begin{figure}
\begin{centering}
\epsfig{figure=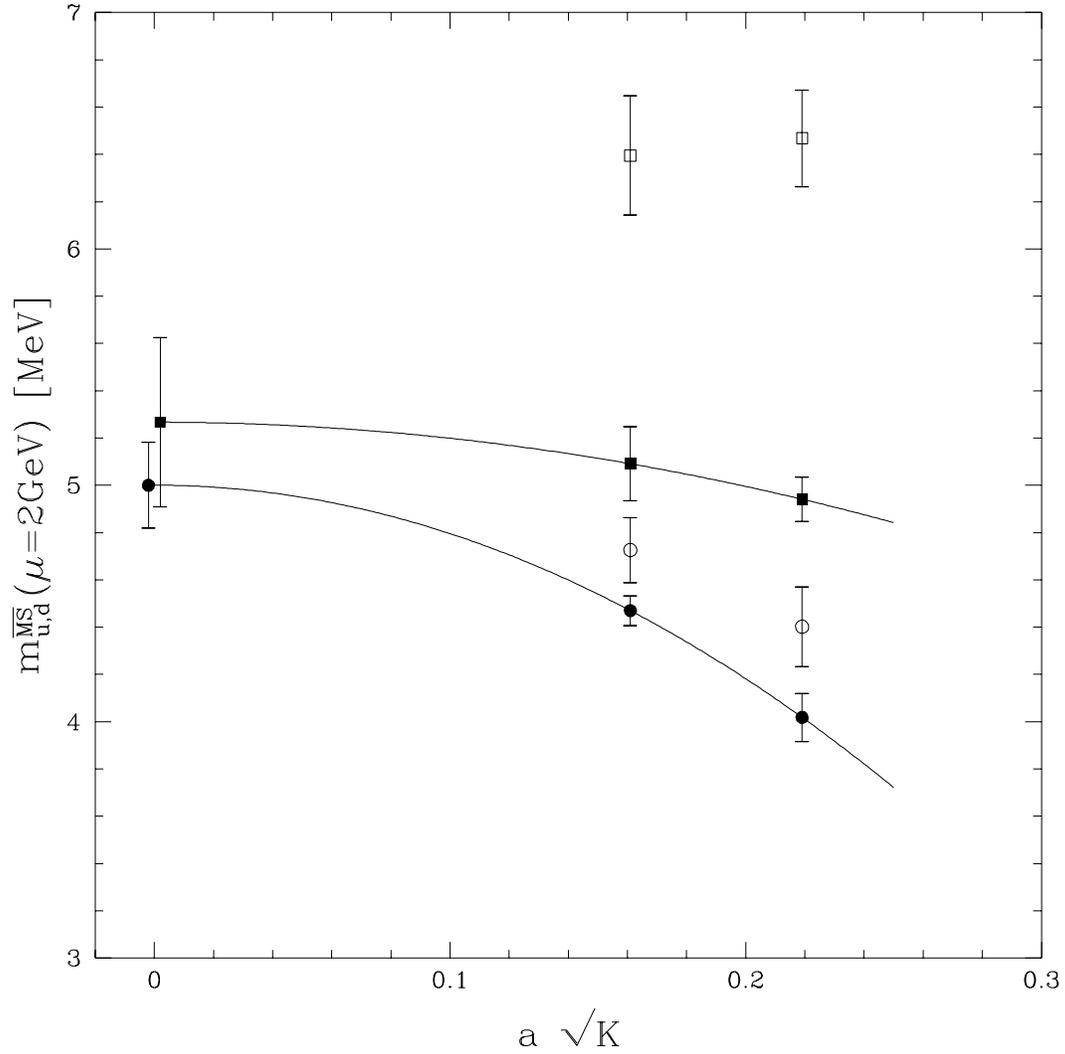,height=15.0cm,width=15.0cm}
\caption{The light quark mass $m_{u,d}^{\overline{MS}}$ as a function of the 
lattice spacing for improved fermions using the Ward identity ($\kreiso$)
and standard method ($\blacksquare$). This is compared with the Wilson result 
for the Ward identity ($\kreisv$) and standard method ($\bbox$).
The curves are quadratic extrapolations to the continuum limit.}
\label{mud}
\end{centering}
\end{figure}

\begin{figure}
\begin{centering}
\epsfig{figure=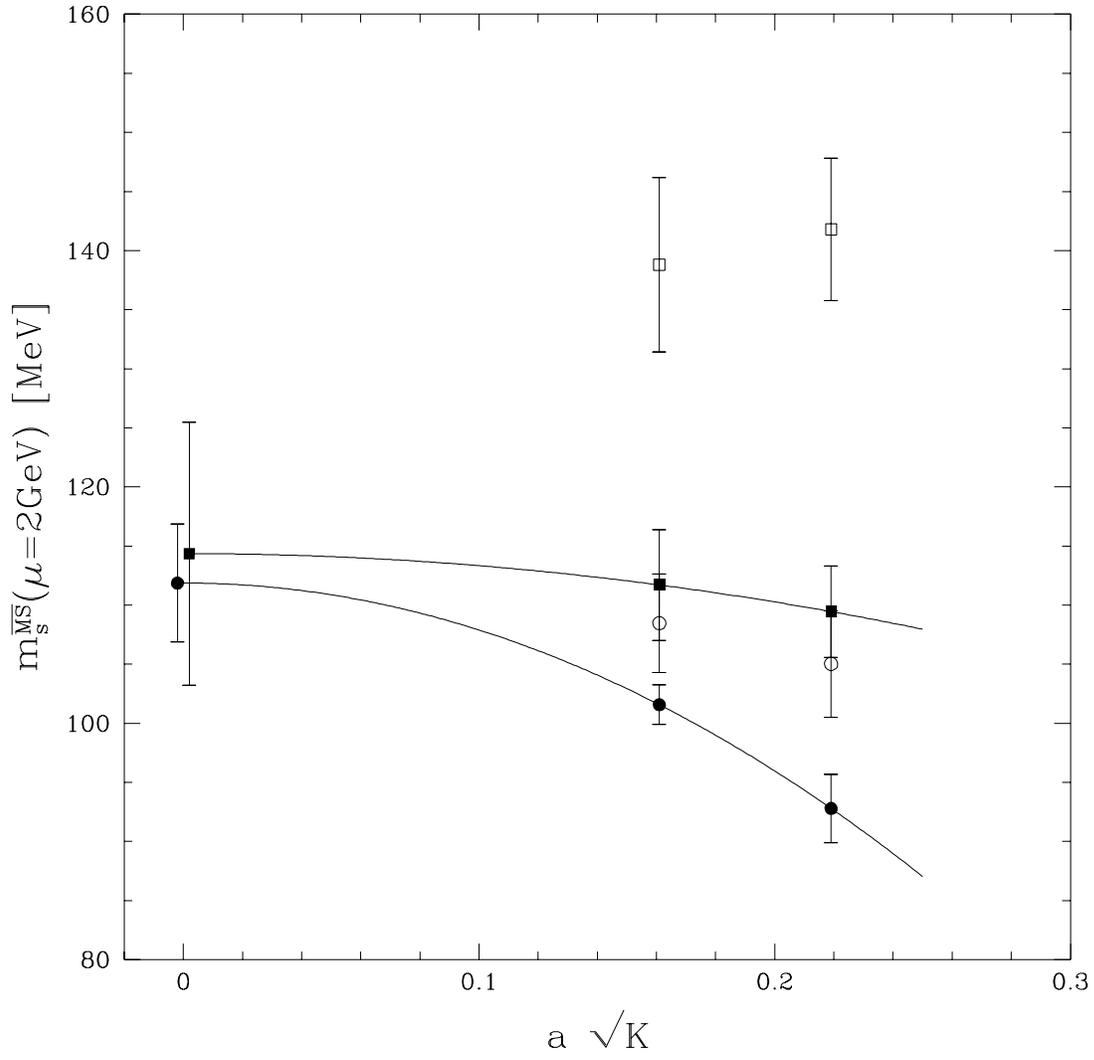,height=15.0cm,width=15.0cm}
\caption{The same as fig.~\protect\ref{mud} but for the strange quark 
mass $m_s^{\overline{MS}}$.}
\label{ms}
\end{centering}
\end{figure}

The results of the Ward identity and the standard method may differ by 
$O(a^2)$ effects, and they do.
We can `fit' the $a$ dependence by
\begin{equation}
m_q^{\overline{MS}} = c_0 + c_2 a^2.
\end{equation}
The result of the fit is shown in figs.~\ref{mud}, ~\ref{ms}. The continuum 
values from this fit are given in table~\ref{mMeV}. We find that the two methods give
consistent results in the continuum limit. Taking the statistical average
of the two results we obtain the continuum values
\begin{eqnarray}
m_{u,d}^{\overline{MS}}(2 \mbox{GeV}) &=& 5.1 \pm 0.2\, \mbox{MeV}, \\
m_s^{\overline{MS}}(2 \mbox{GeV}) &=& 112 \pm 5\, \mbox{MeV}.
\end{eqnarray}
The Ward identity method appears to have larger $O(a^2)$ effects than the
standard method.

We may compare our results with the prediction of chiral perturbation theory,
which cannot give absolute values but can determine the ratio of $m_s$ to
$m_{u,d}$. A recent calculation gives~\cite{leut} $m_s/m_{u,d} = 24.4 \pm 1.5$.
We find $m_s/m_{u,d} = 22.2 \pm 1.2$.

\subsection*{{\it Wilson Fermions}}

Let us now consider the case of Wilson fermions. We proceed in the same way as
before. The situation here is that $Z_A(am)$ is known non-perturbatively 
only for $\beta = 6.0$ \cite{Oetal1}, and that $b_A$, $b_P$ and $b_m$ are
only known to tree-level. So for $Z_A$ we use the tadpole improved 
perturbative value
\begin{equation}
Z_A = \left[ 1 - 
\frac{\alpha_s^{\overline{MS}}(1/a)}{4 \pi} \, 7.901 \right] \, u_0,
\label{zw}
\end{equation} 
and for $b_A$, $b_P$ and $b_m$ we take the tree-level results. 
Comparing $Z_A$ with the non-perturbative determination at 
$\beta = 6.0$~\cite{Oetal1},
as well as with a non-perturbative calculation at $\beta = 5.9$, $6.1$ and 
$6.3$ using  
the Ward identity~\cite{aoki}, we find good agreement.
The renormalization constants $Z_P^{\overline{MS}}(a \mu = 1)$ and 
$Z_m^{\overline{MS}}(a \mu = 1)$ are obtained from 
eqs.~(\ref{zpimp}),~(\ref{zmimp}) by setting $c_{SW} = 0$. The resulting quark
masses are given in table~\ref{mMeV}, and they are plotted and compared
with the improved results in figs.~\ref{mud},~\ref{ms}. 
In this case we expect discretization errors of $O(a)$ instead of $O(a^2)$.
So it is not surprising that the Ward identity and the standard method
give results which are far apart. We find that the Ward identity method
gives mass values which are closer to the continuum result.
                 
\begin{figure}
\begin{centering}
\epsfig{figure=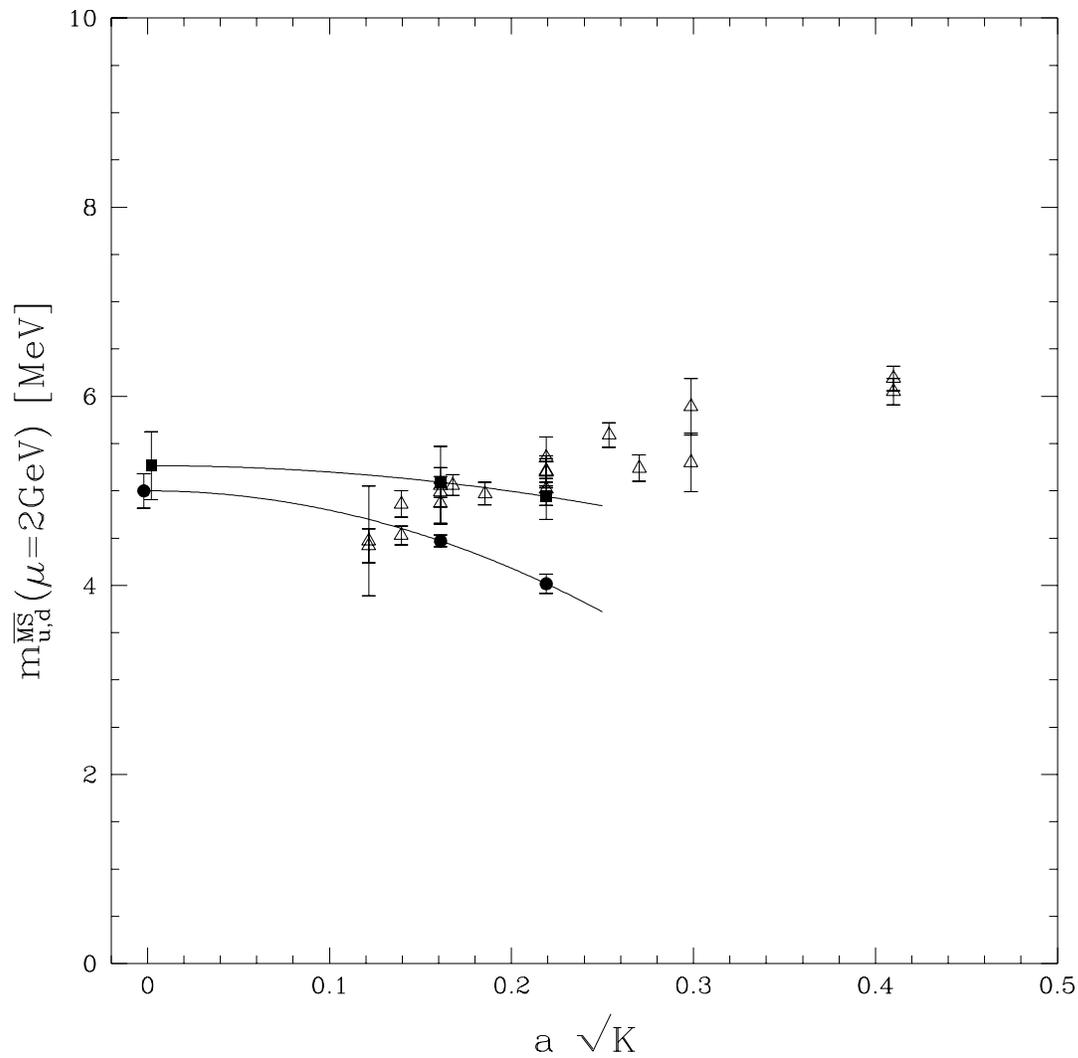,height=15.0cm,width=15.0cm}
\caption{The light quark mass $m_{u,d}^{\overline{MS}}$ for improved fermions
from fig.~\ref{mud} compared with the world Wilson masses ($\triangle$) 
compiled in~\protect\cite{gupta}.}
\label{mudall}
\end{centering}
\end{figure}

\begin{figure}
\begin{centering}
\epsfig{figure=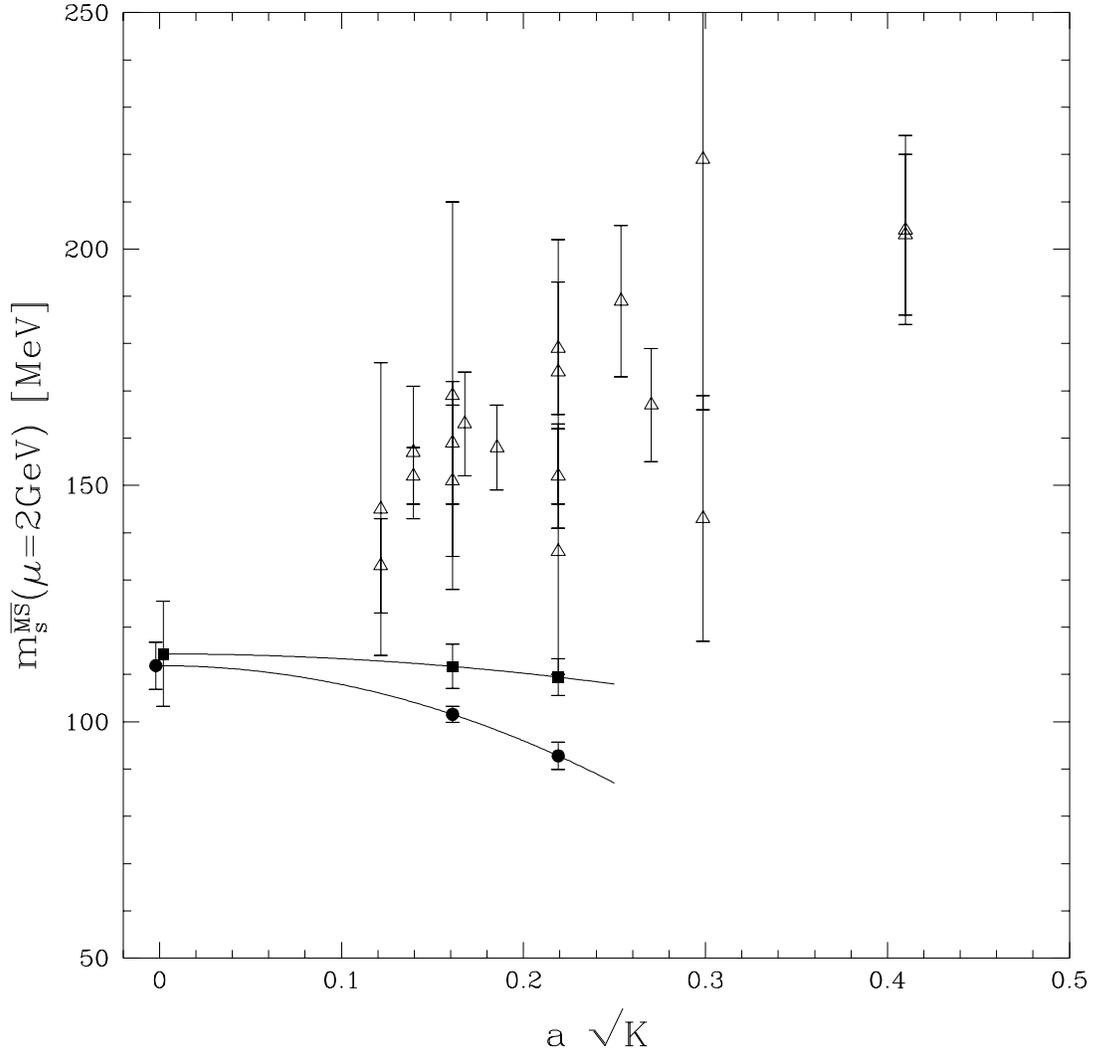,height=15.0cm,width=15.0cm}
\caption{The strange quark mass $m_s^{\overline{MS}}$ for improved fermions
from fig.~\ref{ms} compared with the world Wilson masses ($\triangle$) compiled
in~\protect\cite{gupta}. These authors use the $\phi(1020)$ meson to determine
the strange quark mass.}
\label{msall}
\end{centering}
\end{figure}

Finally, in figs.~\ref{mudall},~\ref{msall} we compare our improved quark
masses with the world data of Wilson quark masses as compiled in 
ref.~\cite{gupta} for the standard method. These authors use the $\rho$ 
mass extrapolated to the chiral limit to set the scale. At $\beta = 6.0$
the scale set by the string tension and by the Wilson action $\rho$ mass 
differ by about 20\% which explains the difference between our Wilson data
and the world data in figs.~\ref{mudall}, \ref{msall}. We see that
the improved action improves the scaling behavior.

\section{Decay Constants}
\label{decay}

The pion decay constant $f_\pi$ is well known experimentally and can be 
determined from the two-point correlation functions on the lattice as well, 
allowing for a
further test of scaling of the improved theory.
We shall also look at the decay constants of the $K$, $\rho$, $K^*$ and 
the $a_1$ meson. 

In Euclidean space at zero three-momentum we define
\begin{eqnarray}
   \langle 0| {\cal A}_4 |\pi\rangle &=& m_\pi f_\pi,   
                               \nonumber \\
   \langle 0| {\cal A}_i |a_1,\lambda\rangle
              &=& e(\lambda)_i\, m_{a_1}^2 f_{a_1},
                            \\
   \langle 0| {\cal V}_i |\rho,\lambda\rangle
              &=& e(\lambda)_i\frac{m_\rho^2}{f_\rho}, \nonumber
   \label{ops}  
\end{eqnarray}
where ${\cal A}$ and ${\cal V}$ are the renormalized axial vector and vector
current, respectively, and $e(\lambda)$ is the polarization vector with
$\sum_\lambda e^*_i(\lambda) e_j(\lambda) = \delta_{i j}$. The pseudoscalar
and vector states are normalized by
\begin{equation}
\langle p|p' \rangle = (2\pi)^3 2 p_0 \delta(\vec{p} - \vec{p}\,').
\end{equation}
Note that our $f_{a_1}$ is defined to be dimensionless.

In the improved theory the renormalized operators are
\begin{eqnarray}
   {\cal A}_\mu &=& (1 + b_A am) Z_A (A_\mu + c_A a\partial_\mu P), \\
   {\cal V}_\mu &=& (1 + b_V am) Z_V (V_\mu + 
\mbox{i} c_V a\partial_\lambda T_{\mu\lambda}),
\end{eqnarray}
where $V_\mu = \bar{\psi}\gamma_\mu\psi$ and
$T_{\mu\nu} = \bar{\psi}\sigma_{\mu\nu}\psi$ are the vector and tensor
operators, respectively.
We use the definition $\sigma_{\mu\nu}=\mbox{i}[\gamma_\mu,\gamma_\nu]/2$. 
Both currents are (partially) conserved, and hence no scale enters into their
definition.
The renormalization constant $Z_A$ and the improvement
coefficients $c_A$ and $b_A$ have already been given in the last section. 
The renormalization constant $Z_V$ and the coefficients $b_V$ and $c_V$ have 
been computed non-perturbatively in ref.~\cite{Letal4,Sommer2}. 
At $\beta = 6.0$ the values are $Z_V = 0.7780$, $b_V = 1.472$ and 
$c_V = - 0.32(6)$, and at $\beta = 6.2$ the numbers are $Z_V = 0.7927$, 
$b_V = 1.409$ and $c_V = - 0.22(7)$. 
While for most of these quantities the authors have given fit formulae 
in $g^2$, for $c_V$ we have read the numbers from the graph in~\cite{Sommer2},
as no such formula exists yet.
We have also determined $Z_V$ and $b_V$ at
$\beta = 6.0$ from our nucleon three-point functions and find consistent
results.

On the lattice we extract the meson decay constant from 
two-point correlation functions.
For large times we expect that
\begin{eqnarray}
   C_{{\cal O}_1 {\cal O}_2}(t)
      &=& \langle {\cal O}_1(t) {\cal O}_2^\dagger(0) \rangle
                              \nonumber  \\
      &=&   \frac{1}{2m_H}
           \left[ \langle 0|{\cal O}_1 |H\rangle
               \langle 0|{\cal O}_2 |H\rangle^* e^{-m_H t} +
               \langle 0|{\cal O}_1^\dagger |H\rangle^*
               \langle 0|{\cal O}_2^\dagger |H\rangle e^{-m_H (T-t)}
           \right]
                                \\
      &\equiv& 
           A_{{\cal O}_1 {\cal O}_2}
           \left[ e^{-m_H t} +
                  \eta_1 \eta_2 e^{-m_H (T-t)} \right], \nonumber
\label{2ptfun}
\end{eqnarray}
where 
${\cal O}(t)$ is of the form 
$V_s^{-\frac{1}{2}}\sum_{\vec{x}} \bar{\psi}(\vec{x},t)
\Gamma \psi(\vec{x},t)$, $V_s$ being the spatial volume of the lattice,
and ${\cal O}^\dagger = \eta {\cal O}$ with 
$\eta = \pm 1$ being given by 
$\gamma_4 \Gamma^\dagger \gamma_4 = \eta \Gamma$. 
The $\eta$ factor tells us how ${\cal O}$ behaves under time reversal, i.e.
whether the two-point
function is symmetric or antisymmetric with respect to $t \rightarrow T-t$.
Here $T$ is the temporal extent of the lattice.  
In general we have computed correlation functions with
local ($L$) and smeared ($S$) operators.

We shall now consider the appropriate matrix elements separately.
We start with those matrix elements necessary for the $\pi$.
With our conventions we set
\begin{eqnarray}
   \langle 0|A_4|\pi\rangle
          &=& m_\pi f_\pi^{(0)},
                               \nonumber \\[-0.2em]
          & & \\
   \langle 0|a \partial_4 P|\pi\rangle 
          &=& - \sinh am_\pi \langle 0|P|\pi\rangle
                       = m_\pi a f_\pi^{(1)}, \nonumber
\label{fpi1}
\end{eqnarray}
where $f^{(0)}$, $f^{(1)}$ are defined to be real and positive.
By computing $C_{A_4P}^{LS}$ and
$C_{PP}^{SS}$ 
we find for the matrix element of $A_4$ from eq.~(\ref{2ptfun}) 
\begin{equation}
   m_\pi f_\pi^{(0)}  =  - 2\kappa \frac{\sqrt{2 m_\pi}
                       A^{LS}_{A_4P}}{\sqrt{A^{SS}_{PP}}},
\end{equation}
and for the matrix element of $\partial_4P$ we obtain from the ratio
of the $C_{PP}^{LS}$ and $C_{A_4P}^{LS}$ correlation functions
\begin{equation}
   \frac{a f_\pi^{(1)}}{f_\pi^{(0)}} =
                \sinh am_\pi \frac{A^{LS}_{PP}}{A^{LS}_{A_4P}}.
\label{fpi2}
\end{equation}
Alternatively, we can take the time derivative from the plateau
in the correlation function. Numerically we found that it made very little
difference to the result.

For the $a_1$ we set
\begin{equation}
   \langle 0|A_i|a_1,\lambda\rangle
                    = e(\lambda)_i\, m_{a_1}^2 f_{a_1}^{(0)},
\label{fa11}
\end{equation}
and we find
\begin{equation}
    m_{a_1}^2 f_{a_1}^{(0)}
        = 2\kappa \frac{\sqrt{2m_{a_1}} \sum_k A_{A_kA_k}^{LS}}
           {\sqrt{3 \sum_k A_{A_kA_k}^{SS}}}.
\end{equation}
For the $\rho$ we set
\begin{eqnarray}
   \langle 0|V_i|\rho,\lambda\rangle 
            &=& e(\lambda)_i\, m_\rho^2 f_\rho^{(0)},
                               \nonumber \\
            & & \\[-0.2em]
   \langle 0|a \partial_4 T_{i4}|\rho,\lambda\rangle 
            &=&
                -\sinh am_\rho 
                 \langle 0|T_{i4}|\rho,\lambda\rangle 
            = \mbox{i} e(\lambda)_i\, m_\rho^2 a f_\rho^{(1)}, \nonumber
\label{frho1}
\end{eqnarray}
and we obtain
\begin{equation}
    m_\rho^2 f_\rho^{(0)}
        =  2\kappa \frac{\sqrt{2m_\rho} \sum_k A_{V_kV_k}^{LS}}
             {\sqrt{3 \sum_k A_{V_kV_k}^{SS}}}
\end{equation}
and 
\begin{equation}
   \frac{a f_\rho^{(1)}}{f_\rho^{(0)}}
       = - \mbox{i} \sinh am_\rho
   \frac{\sum_k A_{T_{k4}T_{k4}}^{LS} \sqrt{\sum_k A_{V_{k}V_{k}}^{SS}}}
              {\sum_k A_{V_kV_k}^{LS} \sqrt{\sum_k A_{T_{k4}T_{k4}}^{SS}}}.
\end{equation}
          
In tables~\ref{decay60} and \ref{decay62} we give
the lattice results for the matrix elements calculated from the
above formulas. The fits to the correlation functions, as
for the masses, are all made using the bootstrap method.

Collecting all the terms, the physical decay constants are given by
\begin{eqnarray}
f_\pi &=& (1 + b_A am) Z_A (f_\pi^{(0)} + c_A a f_\pi^{(1)}), \nonumber \\
f_{a_1} &=& (1 + b_A am) Z_A f_{a_1}^{(0)}, \label{fphys} \\
1/f_\rho &=& (1 + b_V am) Z_V (f_\rho^{(0)} + c_V a f_\rho^{(1)}).
\nonumber
\end{eqnarray}
When the improvement terms are weighted with the appropriate $c$ factors,
they contribute about 10-20\% at $\beta = 6.0$ and up to 10\% at $\beta = 6.2$.
It is thus important to improve the operators as well.

\begin{figure}
\begin{centering}
\epsfig{figure=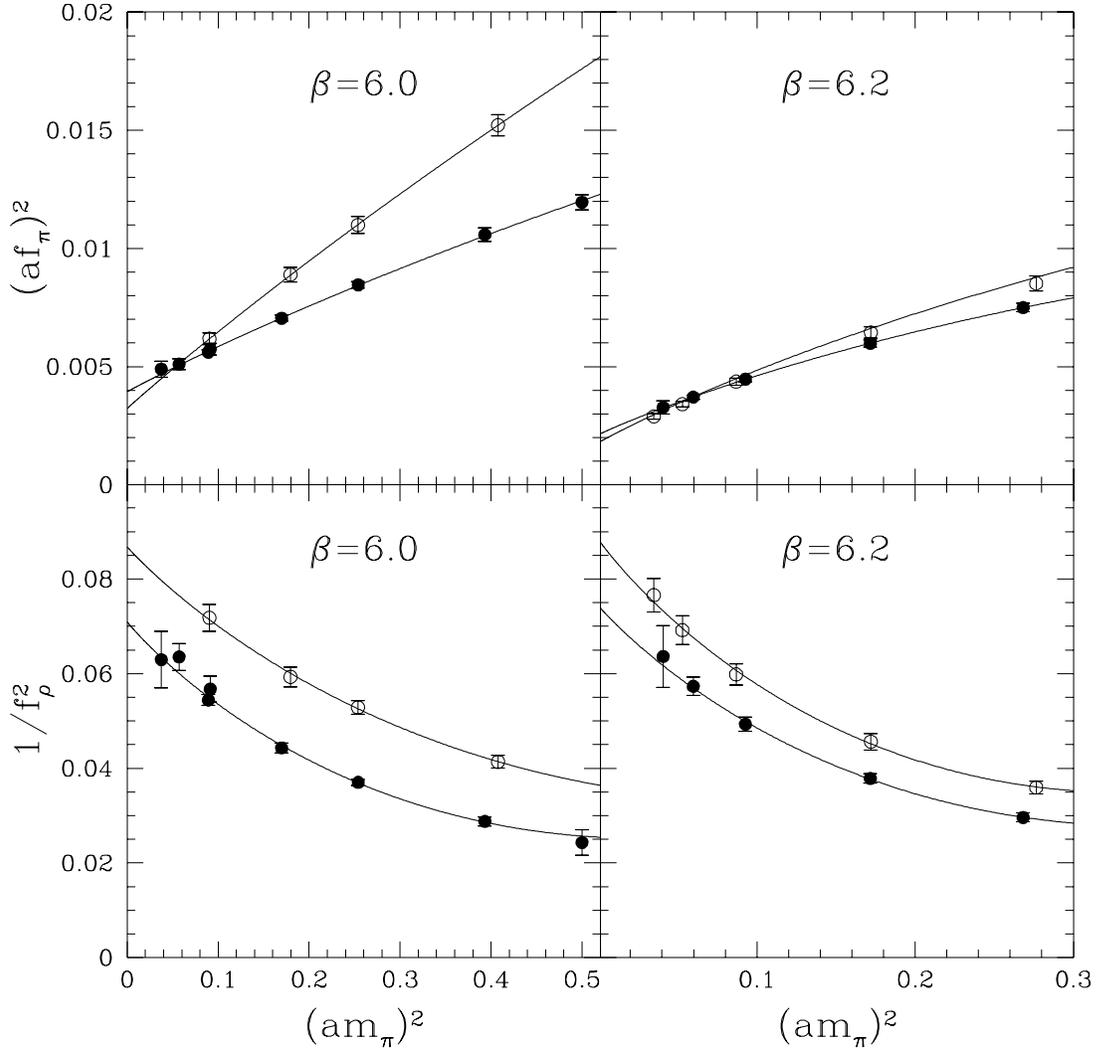,height=15.0cm,width=15.0cm}
\caption{Fits and chiral extrapolations of the decay constants $f_\pi$ and 
$f_\rho$ for improved ($\kreiso$) and Wilson ($\kreisv$) fermions.}
\label{fmass}
\end{centering}
\end{figure}

To perform the chiral extrapolation, we make fits similar to those for the 
hadron masses, namely
\begin{eqnarray}
f_\pi^2 &=& b_0 + b_2 m_\pi^2 + b_3 m_\pi^3, \label{pi} \\
f_{a_1}^2 &=& b_0 + b_2 m_\pi^2 + b_3 m_\pi^3, \\
1/f_\rho^2 &=& b_0 + b_2 m_\pi^2 + b_3 m_\pi^3.
\label{rho}
\end{eqnarray}
We decided to fit the square of the decay constants rather than the decay
constants themselves because this shows less curvature. 
The fits and the data are shown in fig.~\ref{fmass} for 
$f_\pi$ and $f_\rho$. We compare this result with the meson decay constants 
computed with the Wilson action. These follow from eq.~(\ref{fphys}) with
$c_A, c_V = 0$. For $Z_A$ we use the tadpole improved value given in 
eq.~(\ref{zw}), and for $b_A$ we take the tree-level result ($b_A = 1$). The 
renormalization constant $Z_V$ (in the chiral limit) has been determined
non-perturbatively from a two-point correlation function of the local vector
current~\cite{aoki} at $\beta = 5.9$, $6.1$ and $6.3$. 
Unlike the case of $Z_A$, we find significant differences between this
determination and our determination using the nucleon three-point function.
The latter gives $Z_V = 0.651(15)$ at $\beta = 6.0$ which is close to the
tadpole improved result. This indicates large $O(a)$ effects. Since we
are applying $Z_V$ to a two-point function, we chose to use the 
non-perturbative result from ref.~\cite{aoki}.
We interpolate this 
result to $\beta = 6.0$ and $6.2$ and find $Z_V = 0.565$ and $0.618$,
respectively. For $b_V$ we again take the tree-level result. 
Although the individual 
contributions of the improvement terms are significant, the overall result 
for $f_\pi$ in fig. \ref{fmass} is not much changed when compared with the 
Wilson case for smaller 
quark masses. For larger quark masses, especially at $\beta = 6.0$, the
Wilson $f_\pi$ is larger.
The situation is different for $f_\rho$. Here we find a 
systematic difference of 10-20\% at $\beta = 6.0$ and approximately 10\%
at $\beta = 6.2$ for all quark masses. In both cases the difference between 
the two actions becomes smaller with increasing $\beta$ as one would expect.

Our results extrapolated to the chiral limit are given in table~\ref{decaychi},
and we compare $f_\pi$ and $f_\rho$ with experiment in fig. 15. 
For $f_\pi$ we find reasonable agreement of the
improved results with the experimental value using, as before, the string
tension as the scale. 
When including the data of ref.~\cite{wein}, one sees that the Wilson 
results lie
lower, and it appears that the values are increasing as we approach the
continuum limit.
For $f_\rho$ both our improved and Wilson results lie within 5\% of the
experimental value. There is, however, a definite difference as we 
previously remarked. The Wilson numbers lie above the experimental value, 
while the improved ones lie below. One must remember though that in the 
Wilson case there is a systematic error in the renormalization constant
$Z_V$ which may be larger than the statistical errors in the figure.
The experimental number for the decay constant of the $a_1$ is~\cite{wingate} 
$f_{a_1} = 0.17(2)$ (in our notation). The agreement between experimental 
and lattice values is encouraging.

We can avoid errors from extrapolating to the chiral limit by considering
quark masses within our data range, as we have already done in 
figs.~\ref{scalingplot} and \ref{rhopiplot}. The most physical $\kappa$ 
values to use
are those corresponding to the $K$ mass. To obtain the decay
constants we take eqs.~(\ref{pi}) and (\ref{rho}) at $m_\pi = m_K$.
(Remember that we are using $m_\pi$ as a generic name for the pseudoscalar
meson mass.) We give the results for $f_K$ and $f_{K^*}$ in 
table \ref{decaychi}, and 
in fig. 16 we show the scaling behavior together with the experimental value
for $f_K$.
We find the errors to be substantially reduced. For $f_K$ we see no difference
between improved and Wilson results, both lying 10\% below the
experimental value. For $f_{K^*}$ the error bars have become small enough 
to attempt an extrapolation to the continuum limit. The curves are a
simultaneous fit, linear for the Wilson and quadratic for the improved data,
constrained to agree in the continuum limit. In this quantity there appear to
be large $O(a^2)$ effects in the improved case.

\section{Conclusions}

The goal of this paper was to investigate the scaling behavior of $O(a)$
improved fermions. If scaling is good, the results we get should already be
close to the continuum values for present values of the coupling. To this end
we have done simulations for two values of $\beta$ and looked at two-point 
correlation functions from which we derive hadron masses, quark masses and
meson decay constants.

First we looked at hadron masses. The most visible difference between Wilson
and improved fermions is that the $\rho$ mass is much lighter in the Wilson 
case at comparable pion masses. In fig.~\ref{scalingplot} we see that the
improved action 
has brought the $\rho$ mass closer to its physical value when we
use the string tension to set the scale. In this figure we have compared the
Wilson action $\rho$ masses at many different scales. We see a linear behavior
in the lattice spacing $a$ as one would expect. For improved fermions we find
the discretization errors reduced for our couplings.

A problem with Wilson fermions was that they could not describe the
vector-pseudoscalar mass splitting adequately. This problem seems to be
cured by using improved fermions.

Quark masses are important parameters in the Standard Model. Experimentally, 
their values are poorly known, and a reliable lattice determination would be 
useful. Using two different methods, we have determined the light and strange
quark masses. Our results can be seen in figs.~\ref{mud},~\ref{ms}. Both 
methods
give consistent results for improved fermions. In the continuum limit we
find for the average of $u$ and $d$ quark masses
$m_{u,d}^{\overline{MS}}(2\,\mbox{GeV}) = 5.1 \pm 0.2\,\mbox{MeV}$ and 
$m_s^{\overline{MS}}(2\,\mbox{GeV}) = 112 \pm 5\,\mbox{MeV}$. In
the Wilson case the discrepancy between the two methods is much larger, hinting
at substantial $O(a)$ effects.

When calculating the decay constants, an advantage of using the improved
theory is that the renormalization constants and improvement coefficients
for $f_\pi$, $f_{a_1}$, $f_\rho$ and $f_{K^*}$ are known. For $f_K$ we
still have to use the perturbative values of $b_A$ because they have not yet
been computed non-perturbatively. A systematic uncertainty in the Wilson
case lies in the choice of the renormalization constants. While the results
are in reasonable agreement with phenomenology, the data are at present not
precise enough to discuss an extrapolation to the continuum limit, with the
possible exception of $f_{K^*}$. In that case it looks that there are 
relatively large $O(a^2)$ effects between $\beta = 6.0$ and $6.2$.

Our general conclusion is that the Wilson action at $\beta = 6.0$ has
$O(a)$ errors of up to 20\% compared to the continuum extrapolation. The 
non-perturbatively $O(a)$ improved theory still shows $O(a^2)$ effects of
up to 10\% at $\beta = 6.0$, except for the Ward identity quark masses
where the effect is somewhat larger. 
If one wants to go to smaller values of $\beta$, 
one probably will have to reduce the $O(a^2)$ errors as well. Going to 
$\beta = 6.2$ reduces $a^2$ by a factor of almost two, bringing
discretization errors down to 5\% or less. To achieve a one percent accuracy 
would require calculations at several $\beta$ values and an 
extrapolation to $a = 0$.

\section*{Acknowledgement}

This work was supported in part by the Deutsche Forschungsgemeinschaft. The
numerical calculations were performed on the Quadrics computers at 
DESY-Zeuthen. We wish to thank the operating staff for their support.
We furthermore thank Hartmut Wittig for help with table \ref{stringdata}
and Henning Hoeber for communicating his new string tension results to us
prior to publication.

\clearpage
\section*{Tables}

\begin{table}[h]
\vspace*{2.5cm}
\begin{center}
\begin{tabular}{|c|l|l|l|l|l|l|l|} \hline
\multicolumn{8}{|c|}{$\beta = 5.7$} \\ \hline
\multicolumn{8}{c}{ } \\[-0.8em] \hline
\multicolumn{8}{|c|}{$c_{SW} = 1.0$} \\ \hline
$V$ & \multicolumn{1}{c|}{$\kappa$} & \multicolumn{1}{c|}{$am_\pi$} &  
\multicolumn{1}{c|}{$am_\rho$} & \multicolumn{1}{c|}{$am_N$} &
\multicolumn{1}{c|}{$am_{a_0}$} & \multicolumn{1}{c|}{$am_{a_1}$} &
\multicolumn{1}{c|}{$am_{b_1}$} \\ \hline 
    & 0.1500 & 0.5028(17) & 0.757(7) & 1.135(18) & 1.36(10) & 1.61(19) & 
1.06(16)\\
$16^3 32$ & 0.1510 & 0.414(2) & 0.711(8) & 1.040(17) & 1.11(17) & 1.31(12) & 
1.11(17) \\
    & 0.1520 & 0.288(5) & 0.660(19) & 0.92(3) & $-$ & 1.09(17) & 
1.25(20) \\ \hline
\multicolumn{1}{|c|}{$c.~l.$} & 0.15280(14) & 0 & {\it 0.605(24)} & 
{\it 0.797(49)} & $-$ & {\it 0.70(36)} & {\it 1.31(30)} \\ \hline
\multicolumn{8}{c}{ } \\[-0.8em] \hline
\multicolumn{8}{|c|}{$c_{SW} = 2.25$} 
\\ \hline
$V$ & \multicolumn{1}{c|}{$\kappa$} & \multicolumn{1}{c|}{$am_\pi$} &  
\multicolumn{1}{c|}{$am_\rho$} & \multicolumn{1}{c|}{$am_N$} &
\multicolumn{1}{c|}{$am_{a_0}$} & \multicolumn{1}{c|}{$am_{a_1}$} &
\multicolumn{1}{c|}{$am_{b_1}$} \\ \hline 
    & 0.1270 & 0.841(3) & 1.087(8) & 1.588(12) & 1.64(10) & 1.56(7) & 
1.48(13) \\ 
    & 0.1275 & 0.791(4) & 1.053(10) & 1.518(23) & 1.56(7) & 1.51(5) & 
1.53(9)\\
    & 0.1280 & 0.736(3) & 1.022(11) & 1.453(18) & 1.50(7) & 1.46(4) & 
1.42(7) \\[-0.6em]
$16^3 32$ & & & & & & & \\[-0.6em]
    & 0.1285 & 0.672(5) & 0.988(9) & 1.399(24) & 1.57(14) & 1.41(6) & 
1.39(7) \\ 
    & 0.1290 & 0.607(7) & 0.955(8) & 1.320(20) & 1.59(14) & 1.34(8) & 
1.33(11) \\
     & 0.1295 & 0.519(11) & 0.922(16) & 1.23(3) & $-$ & 1.28(10) & 
1.33(16) \\ \hline
\multicolumn{1}{|c|}{$c.~l.$} & 0.13074(29) & 0 & {\it 0.793(19)} & 
{\it 0.948(46)} &  
{\it 1.43(33)} & {\it 1.06(16)} & {\it 1.13(25)} \\ \hline
\end{tabular}
\end{center}
\vspace*{0.3cm}
\caption{Hadron masses at $\beta = 5.7$ for  
Sheikholeslami-Wohlert fermions with
$c_{SW} = 1$ and $2.25$. 
In the bottom row we give $\kappa_c$ and the mass values extrapolated to the 
chiral limit. The numbers in roman ({\it italic}) are from three-parameter 
(two-parameter) fits. The errors are bootstrap errors.}
\label{tm57}
\end{table}

\clearpage
\begin{table}
\begin{center}
\begin{tabular}{|c|l|l|l|l|l|l|l|} \hline
\multicolumn{8}{|c|}{$\beta = 6.0$} \\ \hline
\multicolumn{8}{c}{ } \\[-0.8em] \hline
\multicolumn{8}{|c|}{$c_{SW} = 0$} \\ \hline
$V$ & \multicolumn{1}{c|}{$\kappa$} & \multicolumn{1}{c|}{$am_\pi$} &  
\multicolumn{1}{c|}{$am_\rho$} & \multicolumn{1}{c|}{$am_N$} &
\multicolumn{1}{c|}{$am_{a_0}$} & \multicolumn{1}{c|}{$am_{a_1}$} &
\multicolumn{1}{c|}{$am_{b_1}$} \\ \hline 
    & 0.1487 & 0.6384(18) \hspace*{-0.25cm}& 0.683(2) & 1.071(7) & 0.885(19) & 0.933(13) & 
0.940(19) \\ 
    & 0.1515 & 0.5037(8) & 0.5696(10) \hspace*{-0.25cm}& 0.9019(17) \hspace*{-0.25cm}& 0.817(7) & 0.851(7) & 
0.849(13)\\[-0.6em]  
$16^3 32$ & & & & & & & \\[-0.6em]
    & 0.1530 & 0.4237(8) & 0.5080(11) \hspace*{-0.25cm}& 0.7977(20) \hspace*{-0.25cm}& 0.763(11) & 0.797(6) & 
0.809(7) \\
    & 0.1550 & 0.3009(10) \hspace*{-0.25cm}& 0.4264(14) \hspace*{-0.25cm}& 0.6517(30) \hspace*{-0.25cm}& 0.735(15) & 0.717(12) & 
0.736(9) \\ \hline
    & 0.1550 & 0.292(2) & 0.418(5) & 0.638(8) & 0.610(48) & 0.657(33) & 
0.659(35) \\
$24^3 32$ & 0.1558 & 0.229(2) & 0.384(7) & 0.555(12) & 0.616(90) & 0.613(41) & 
0.638(38)  \\ 
    & 0.1563 & 0.179(3) & 0.358(11) & 0.488(22) & 0.88(15) & 0.584(52) & 
0.615(44) \\ \hline 
\multicolumn{1}{|c|}{$c.~l.$} & 0.15713(3) \hspace*{-0.25cm}& 0 & 0.327(6) & 0.412(16) &  
{\it 0.658(19)} \hspace*{-0.25cm}& {\it 0.632(14)} \hspace*{-0.25cm}& {\it 0.650(13)} \hspace*{-0.25cm}\\ \hline
\multicolumn{8}{c}{ }\\[-0.8em] \hline
\multicolumn{8}{|c|}{$c_{SW} = 1.769$}\\ \hline
$V$ & \multicolumn{1}{c|}{$\kappa$} & \multicolumn{1}{c|}{$am_\pi$} &  
\multicolumn{1}{c|}{$am_\rho$} & \multicolumn{1}{c|}{$am_N$} &
\multicolumn{1}{c|}{$am_{a_0}$} & \multicolumn{1}{c|}{$am_{a_1}$} &
\multicolumn{1}{c|}{$am_{b_1}$} \\ \hline 
    & 0.1300 & 0.707(2) & 0.783(6) & 1.190(6) &  &  &   \\ 
    & 0.1310 & 0.627(2) & 0.714(3) & 1.079(7) &  &  &   \\ 
    & 0.1320 & 0.545(5) & 0.644(8) & 0.974(16) &  &  &  \\[-0.6em]
$16^3 32$ & & & & & & & \\[-0.6em] 
    & 0.1324 & 0.5039(7) & 0.6157(16) \hspace*{-0.25cm}& 0.932(4) & 0.779(14) & 0.829(12) & 
0.853(7)\\ 
    & 0.1333 & 0.4122(8) & 0.5502(23) \hspace*{-0.25cm}& 0.821(5) & 0.738(15) & 0.773(7) & 
0.799(10) \\
    & 0.1342 & 0.2988(17) \hspace*{-0.25cm}& 0.487(3) & 0.705(9) & 0.92(5) & 0.68(2) & 
0.775(15) \\ \hline
    & 0.1342 & 0.3020(11) \hspace*{-0.25cm}& 0.491(3) & 0.686(7) & 0.82(3) & 0.715(19) & 
0.758(16) \\
$24^3 32$ & 0.1346 & 0.2388(14) \hspace*{-0.25cm}& 0.467(6) & 0.626(10) & 1.00(8) & 0.684(26) & 
0.745(20)  \\ 
    & 0.1348 & 0.194(4) & 0.448(13) & 0.593(19) & 1.52(20) & 0.664(34) & 
0.736(29) \\ \hline 
\multicolumn{1}{|c|}{$c.~l.$} & 0.13531(1) \hspace*{-0.25cm}& 0 & 0.417(7) & 0.511(15) &  
{\it 0.816(33)} \hspace*{-0.25cm}& {\it 0.625(19)} \hspace*{-0.25cm}& {\it 0.710(14)} \hspace*{-0.25cm}\\ \hline
\end{tabular}
\end{center}
\vspace*{0.3cm}
\caption{Hadron masses at $\beta = 6.0$ for Wilson 
fermions ($c_{SW} = 0$) and improved fermions ($c_{SW} = 1.769$). 
Otherwise the notation is the same as in
table~\protect\ref{tm57}.} 
\label{tm60}
\end{table}

\clearpage
\begin{table}
\begin{center}
\begin{tabular}{|c|l|l|l|l|l|l|l|} \hline
\multicolumn{8}{|c|}{$\beta = 6.2$} \\ \hline
\multicolumn{8}{c}{ } \\[-0.8em] \hline
\multicolumn{8}{|c|}{$c_{SW} = 0$} \\ \hline
$V$ & \multicolumn{1}{c|}{$\kappa$} & \multicolumn{1}{c|}{$am_\pi$} &  
\multicolumn{1}{c|}{$am_\rho$} & \multicolumn{1}{c|}{$am_N$} &
\multicolumn{1}{c|}{$am_{a_0}$} & \multicolumn{1}{c|}{$am_{a_1}$} &
\multicolumn{1}{c|}{$am_{b_1}$} \\ \hline 
    & 0.1468 & 0.5258(12) \hspace*{-0.25cm}& 0.5585(16) \hspace*{-0.25cm}& 0.872(5) & 0.685(8) & 0.700(21) & 
0.695(21)\\ 
    & 0.1489 & 0.4148(13) \hspace*{-0.25cm}& 0.4615(19) \hspace*{-0.25cm}& 0.720(6) & 0.589(8) & 0.624(9) & 
0.626(9) \\
$24^3 48$ & 0.1509 & 0.2947(14) \hspace*{-0.25cm}& 0.3672(27) \hspace*{-0.25cm}& 0.560(10) \hspace*{-0.25cm}& 0.507(14) \hspace*{-0.25cm}& 0.536(13) & 
0.540(13) \\ 
    & 0.1518 & 0.2299(15) \hspace*{-0.25cm}& 0.326(4) & 0.487(12) \hspace*{-0.25cm}& 0.474(20) \hspace*{-0.25cm}& 0.509(16) & 
0.519(17) \\
    & 0.1523 & 0.1867(17) \hspace*{-0.25cm}& 0.307(6) & 0.448(14) \hspace*{-0.25cm}& 0.479(30) \hspace*{-0.25cm}& 0.492(17) & 
0.511(21) \\ \hline 
\multicolumn{1}{|c|}{$c.~l.$} & 0.15336(4) \hspace*{-0.25cm}& 0 & 0.255(9) & 0.342(28) \hspace*{-0.25cm}&  
{\it 0.407(17)} \hspace*{-0.25cm}& {\it 0.449(15)} \hspace*{-0.25cm}& {\it 0.464(16)} \hspace*{-0.25cm}\\ \hline
\multicolumn{8}{c}{ }\\[-0.8em] \hline
\multicolumn{8}{|c|}{$c_{SW} = 1.614$} 
\\ \hline
$V$ & \multicolumn{1}{c|}{$\kappa$} & \multicolumn{1}{c|}{$am_\pi$} &  
\multicolumn{1}{c|}{$am_\rho$} & \multicolumn{1}{c|}{$am_N$} &
\multicolumn{1}{c|}{$am_{a_0}$} & \multicolumn{1}{c|}{$am_{a_1}$} &
\multicolumn{1}{c|}{$am_{b_1}$} \\ \hline  
    & 0.1321 & 0.5179(7) & 0.5738(11) \hspace*{-0.25cm}& 0.877(4) & 0.691(5) & 0.723(6) & 
0.727(6)\\ 
    & 0.1333 & 0.4143(8) & 0.4850(15) \hspace*{-0.25cm}& 0.735(5) & 0.603(10) \hspace*{-0.25cm}& 0.642(5) & 
0.638(8) \\
$24^3 48$ & 0.1344 & 0.3046(9) & 0.4005(26) \hspace*{-0.25cm}& 0.592(9) & 0.532(21) \hspace*{-0.25cm}& 0.563(7) 
& 0.566(9) \\ 
    & 0.1349 & 0.2444(9) & 0.3626(43) \hspace*{-0.25cm}& 0.521(13) \hspace*{-0.25cm}& 0.543(22) \hspace*{-0.25cm}& 0.529(10) & 
0.539(12) \\
    & 0.1352 & 0.2016(11) \hspace*{-0.25cm}& 0.3430(53) \hspace*{-0.25cm}& 0.485(6) & 0.646(53) \hspace*{-0.25cm}& 0.514(13) & 
0.523(25) \\ \hline 
\multicolumn{1}{|c|}{$c.~l.$} & 0.13589(2) \hspace*{-0.25cm}& 0 & 0.287(9) & 0.378(18) \hspace*{-0.25cm}&  
{\it 0.460(21)} \hspace*{-0.25cm}& {\it 0.460(9)} & {\it 0.465(12)} \hspace*{-0.25cm}\\ \hline
\end{tabular}
\end{center}
\vspace*{0.3cm}
\caption{Hadron masses at $\beta = 6.2$ for Wilson 
fermions ($c_{SW} = 0$) and improved fermions ($c_{SW} = 1.614$). 
Otherwise the notation is the same as in table~\protect\ref{tm57}.}
\label{tm62}
\end{table}

\begin{table}[hb]
\begin{center}
\begin{tabular}{|l|l|l|c|l|c|l|c|} \hline
 \multicolumn{1}{|c|}{\raisebox{-1.5ex}{$\beta$}} & 
 \multicolumn{1}{|c|}{\raisebox{-1.5ex}{$c_{SW}$}} & \multicolumn{2}{c|} {eq.~(\ref{linear})} &
 \multicolumn{2}{c|} {eq.~(\ref{chiral})} & 
\multicolumn{2}{c|} {eq.~(\ref{pheno})} \\ \cline{3-8}
 &  & \multicolumn{1}{|c|}{$\kappa_c$} & {$\chi^2/$dof} &
 \multicolumn{1}{|c|}{$\kappa_c$} & {$\chi^2/$dof} &
 \multicolumn{1}{|c|}{$\kappa_c$} & $\chi^2/$dof \\ 
\hline
  & 1.0  & 0.15305(5)& 3.0 & 0.15274(15) & -& 0.15280(14)& - \\[-0.6em]
5.7 &  &  &  & & & & \\[-0.6em]
 & 2.25  & 0.13120(7) & 0.7 & 0.13065(28) & 0.1 & 0.13074(29)& 0.1 \\ 
\hline
 & 0      & 0.15695(1) &17.5& 0.15726(5) &8.6 & 0.15713(3)&6.6 
\\[-0.6em]
6.0 & & & & & & & \\[-0.6em]
 & 1.769  & 0.13521(1) &11.5 & 0.13537(2) &1.5 & 0.13531(1) & 1.0 \\ 
\hline
 & 0      & 0.15308(1) &30.8& 0.15361(8) &0.7 & 0.15336(4) &0.0 
\\[-0.6em]
6.2 & & & & & & & \\[-0.6em]
 & 1.614  & 0.13574(1) &39.6 & 0.13601(3) & 1.2 & 0.13589(2) & 0.1 \\ 
\hline
\end{tabular}
\end{center}
\vspace*{0.3cm}
\caption{The critical values of $\kappa$, $\kappa_c$, of our 
data for the linear (eq.~(\ref{linear})), 
chiral (eq.~(\ref{chiral})) and phenomenological fit 
(eq.~(\ref{pheno})) for
the various $c_{SW}$ parameters.}
\label{kappa}
\end{table}

\clearpage
\begin{table}
\begin{center}
\begin{tabular}{|c|l|l|l|l|l|l|l|} \hline
$\beta$ & \multicolumn{1}{c|}{$\kappa$} & \multicolumn{1}{c|}{$am_\pi$} &  
\multicolumn{1}{c|}{$am_\rho$} & \multicolumn{1}{c|}{$am_N$} &
 \multicolumn{1}{c|}{Lattice}& \multicolumn{1}{c|}{Reference} \\ \hline 
6.30 & 0.1400 & 0.789(4)& 0.804(4)& & $ 32^3 \times 48$
 & \cite{QCDTARO} \\
6.30 & 0.1430 & 0.646(6)& 0.670(5)& & $ 32^3 \times 48$
 & \cite{QCDTARO} \\
6.30 & 0.1460 & 0.4879(12)& 0.5188(18)& 0.8252(42)& $ 24^3 \times 32$
 & \cite{APE} \\
6.30 & 0.1480 & 0.382(4)& 0.429(4)& & $ 32^3 \times 48$
 & \cite{QCDTARO} \\
6.30 & 0.1485 & 0.3480(14)& 0.3990(23)& 0.6340(47)& $ 24^3 \times 32$
 & \cite{APE} \\
6.30 & 0.1498 & 0.2631(19)& 0.3354(30)& 0.5215(67)& $ 24^3 \times 32$
 & \cite{APE} \\
6.30 & 0.1500 & 0.253(6)& 0.333(4)& & $ 32^3 \times 48$
 & \cite{QCDTARO} \\
6.30 & 0.1505 & 0.2093(26)& 0.3012(40)& 0.4506(89)& $ 24^3 \times 32$
 & \cite{APE} \\
 \hline 
6.20 & 0.1468 & 0.5258(12)& 0.5585(16)& 0.872(5)& $ 24^3 \times 48 $
 & this work \\
6.20 & 0.1489 & 0.4148(13)& 0.4615(19)& 0.720(6)& $ 24^3 \times 48$
 & this work \\
6.20 & 0.1509 & 0.2947(14)& 0.3672(27)& 0.560(10)& $ 24^3 \times 48$
 & this work \\
6.20 & 0.1510 & 0.289(1)& 0.366(2)& 0.566(4)& $ 24^3 \times 64$
 & \cite{Rap_mail} \\
6.20 & 0.1515 & 0.254(1)& 0.343(3)& 0.525(6)& $ 24^3 \times 64$
 & \cite{Rap_mail} \\
6.20 & 0.1518 & 0.2299(15)& 0.326(4)& 0.487(12)& $ 24^3 \times 48$
 & this work \\
6.20 & 0.1520 & 0.220(7)& 0.327(9)& 0.495(10)& $ 24^3 \times 48$
 & \cite{UKQCD92} \\
6.20 & 0.1520 & 0.215(1)& 0.321(5)& 0.48(1)& $ 24^3 \times 64$
 & \cite{Rap_mail} \\
6.20 & 0.1523 & 0.1867(17)& 0.307(6)& 0.448(14)& $ 24^3 \times 48$
 & this work \\
6.20 & 0.1526 & 0.158(1)& 0.29(1)& 0.45(3)& $ 24^3 \times 64$
 & \cite{Rap_mail} \\
 \hline 
6.17 & 0.1500 & 0.3866(12)& 0.4458(18)& 0.6966(40)& $32^2 \times 30 \times 40$
 & \cite{Weinga} \\
6.17 & 0.1519 & 0.2631(12)& 0.3572(26)& 0.5460(52)& $32^2 \times 30 \times 40$
 & \cite{Weinga} \\
6.17 & 0.1526 & 0.2064(15)& 0.3245(39)& 0.4848(68)& $32^2 \times 30 \times 40$
 & \cite{Weinga} \\
6.17 & 0.1532 & 0.1455(20)& 0.2965(88)& 0.4097(78)& $32^2 \times 30 \times 40$
 & \cite{Weinga} \\
 \hline 
\end{tabular}
\end{center}
\vspace*{0.3cm}
\caption{World Wilson fermion masses above $\beta = 6.0$.}  
\label{wilsonm1}
\end{table} 

\clearpage
\begin{table}
\begin{center}
\begin{tabular}{|c|l|l|l|l|l|l|l|} \hline
$\beta$ & \multicolumn{1}{c|}{$\kappa$} & \multicolumn{1}{c|}{$am_\pi$} &  
\multicolumn{1}{c|}{$am_\rho$} & \multicolumn{1}{c|}{$am_N$} &
 \multicolumn{1}{c|}{Lattice}& \multicolumn{1}{c|}{Reference} \\ \hline 
6.0 & 0.1450 & 0.8069(7)& 0.8370(9)& 1.3225(28)& $ 24^3 \times 54 $
 & \cite{QCDPAX} \\
6.0 & 0.1487 & 0.6384(18)& 0.683(2)& 1.071(7)& $ 16^3 \times 32$
 & this work \\
6.0 & 0.1515 & 0.5037(8)& 0.5696(10)& 0.9019(17)& $ 16^3 \times 32$
 & this work \\
6.0 & 0.1520 & 0.4772(9)& 0.5486(15)& 0.8669(49)& $ 24^3 \times 54$
 & \cite{QCDPAX} \\
6.0 & 0.1520 & 0.474(1)& 0.545(2)& 0.861(5)& $ 18^3 \times 32$
 & \cite{APE} \\
6.0 & 0.1530 & 0.423(1)& 0.508(3)& 0.801(6)& $ 18^3 \times 64$
 & \cite{Rap_mail} \\
6.0 & 0.1530 & 0.4237(8)& 0.5080(11)& 0.7977(20)& $ 16^3 \times 32$
 & this work \\
6.0 & 0.1530 & 0.422(1)& 0.505(1)& 0.786(3)& $ 32^3 \times 64$
 & \cite{Gupta} \\
6.0 & 0.1540 & 0.364(1)& 0.468(4)& 0.729(7)& $ 18^3 \times 64$
 & \cite{Rap_mail} \\
6.0 & 0.1545 & 0.33076(28)& 0.4425(10)& 0.6777(21)& $ 24^3 \times 64$
 & \cite{JLQCD} \\
6.0 & 0.1550 & 0.298(1)& 0.431(6)& 0.66(1)& $ 18^3 \times 64$
 & \cite{Rap_mail} \\
6.0 & 0.1550 & 0.3009(10)& 0.4264(14)& 0.6517(30)& $ 16^3 \times 32$
 & this work \\
6.0 & 0.1550 & 0.29642(27)& 0.4220(12)& 0.6393(27)& $ 24^3 \times 64$
 & \cite{JLQCD} \\
6.0 & 0.1550 & 0.292(2)& 0.418(5)& 0.638(8)& $ 24^3 \times 32$
 & this work \\
6.0 & 0.1550 & 0.2967(15)& 0.4218(42)& 0.6440(85)& $ 24^3 \times 54 $
 & \cite{QCDPAX} \\
6.0 & 0.1550 & 0.296(1)& 0.422(2)& 0.630(5)& $ 32^3 \times 64$
 & \cite{Gupta} \\
6.0 & 0.1555 & 0.25864(33)& 0.4016(17)& 0.6003(37)& $ 24^3 \times 64$
 & \cite{JLQCD} \\
6.0 & 0.1555 & 0.2588(16)& 0.3982(61)& 0.6007(109)& $ 24^3 \times 54$
 & \cite{QCDPAX} \\
6.0 & 0.1558 & 0.234(1)& 0.387(3)& 0.557(7)& $ 32^3 \times 64$
 & \cite{Gupta} \\
6.0 & 0.1558 & 0.229(2)& 0.384(7)& 0.555(12)& $ 24^3 \times 32$
 & this work \\
6.0 & 0.1563 & 0.1847(27)& 0.353(15)& 0.536(30)& $ 24^3 \times 54 $
 & \cite{QCDPAX} \\
6.0 & 0.1563 & 0.185(1)& 0.361(5)& 0.506(11)& $ 32^3 \times 64$
 & \cite{Gupta} \\
6.0 & 0.1563 & 0.179(3)& 0.358(11)& 0.488(22)& $ 24^3 \times 32$
 & this work \\
 \hline 
\end{tabular}
\end{center}
\vspace*{0.3cm}
\caption{World Wilson fermion masses at $\beta = 6.0$.}
\label{wilsonm2}
\end{table} 

\clearpage
\begin{table}
\begin{center}
\begin{tabular}{|c|l|l|l|l|l|l|l|} \hline
$\beta$ & \multicolumn{1}{c|}{$\kappa$} & \multicolumn{1}{c|}{$am_\pi$} &  
\multicolumn{1}{c|}{$am_\rho$} & \multicolumn{1}{c|}{$am_N$} &
 \multicolumn{1}{c|}{Lattice}& \multicolumn{1}{c|}{Reference} \\ \hline 
5.93 & 0.1543 & 0.4572(26)& 0.5527(40)& 0.8674(102)& $ 24^3 \times 36 $
 & \cite{Weinga} \\
5.93 & 0.1560 & 0.3573(19)& 0.4864(42)& 0.7448(99)& $ 24^3 \times 36$
 & \cite{Weinga} \\
5.93 & 0.1573 & 0.2641(25)& 0.4369(48)& 0.6423(80)& $ 24^3 \times 36$
 & \cite{Weinga} \\
5.93 & 0.1581 & 0.1885(31)& 0.4071(57)& 0.5652(92)& $ 24^3 \times 36$
 & \cite{Weinga} \\
 \hline 
5.85 & 0.1440 & 1.0293(12)& 1.0598(15)& 1.6961(50)& $ 24^3 \times 54$
 & \cite{QCDPAX} \\
5.85 & 0.1540 & 0.6122(11)& 0.6931(27)& 1.1060(55)& $ 24^3 \times 54$
 & \cite{QCDPAX} \\
5.85 & 0.1585 & 0.3761(12)& 0.5294(69)& 0.815(13)& $ 24^3 \times 54$
 & \cite{QCDPAX} \\
5.85 & 0.1585 & 0.378(2)& 0.530(6)& 0.783(10)& $ 16^3 \times 32 $
 & \cite{Bitar} \\
5.85 & 0.1595 & 0.3088(14)& 0.4856(96)& 0.744(17)& $ 24^3 \times 54$
 & \cite{QCDPAX} \\
5.85 & 0.1600 & 0.2730(30)& 0.486(9)& 0.673(9)& $ 16^3 \times 32$
 & \cite{Bitar} \\
5.85 & 0.1605 & 0.2226(21)& 0.434(20)& 0.683(48)& $ 24^3 \times 54$
 & \cite{QCDPAX} \\
 \hline 
5.70 & 0.1600 & 0.6905(31)& 0.8022(56)& 1.3124(135)& $ 24^3 \times 32$
 & \cite{Weinga} \\
5.70 & 0.1600 & 0.6873(24)& 0.8021(29)& 1.2900(60)& $ 16^3 \times 20$
 & \cite{Fukugita} \\
5.70 & 0.1610 & 0.6527(15)& 0.7842(26)& 1.263(5)& $ 12^3 \times 24$
 & \cite{APE} \\
5.70 & 0.1630 & 0.5621(18)& 0.7232(35)& 1.153(6)& $ 12^3 \times 24 $
 & \cite{APE} \\
5.70 & 0.1640 & 0.5080(29)& 0.6822(38)& 1.0738(80)& $ 16^3 \times 20$
 & \cite{Fukugita} \\
5.70 & 0.1650 & 0.4604(22)& 0.6663(45)& 1.039(8)& $ 12^3 \times 24$
 & \cite{APE} \\
5.70 & 0.1650 & 0.4589(22)& 0.6491(73)& 1.0301(104)& $ 24^3 \times 32$
 & \cite{Weinga} \\
5.70 & 0.1663 & 0.3829(26)& 0.6206(103)& 0.9421(131)& $ 24^3 \times 32$
 & \cite{Weinga} \\
5.70 & 0.1665 & 0.3674(39)& 0.6085(58)& 0.915(11)& $ 16^3 \times 20$
 & \cite{Fukugita} \\
5.70 & 0.1670 & 0.3302(30)& 0.6042(83)& 0.919(14)& $ 12^3 \times 24$
 & \cite{APE} \\
5.70 & 0.1675 & 0.2955(24)& 0.5912(125)& 0.8668(177)& $ 24^3 \times 32$
 & \cite{Weinga} \\
 \hline 
\end{tabular}
\end{center}
\vspace*{0.3cm}
\caption{World Wilson fermion masses below $\beta = 6.0$.}
\label{wilsonm3}
\end{table} 

\clearpage
\begin{table}
\begin{center}
\begin{tabular}{|c|l|l|l|l|l|l|l|} \hline
$\beta$ & \multicolumn{1}{c|}{$a\sqrt{K}$} & \multicolumn{1}{c|}{Reference} &  
\multicolumn{1}{c|}{$r_0/a$} & \multicolumn{1}{c|}{Reference} &
\multicolumn{1}{c|}{$r_0 \sqrt{K}$} \\ \hline 
 6.8 & 0.0730(12) & \cite{BaSch} & 16.7(4) &\cite{Bali} & 1.22(4) \\ \hline
 6.5 & 0.1068(10) & \cite{CM_65} & & & \\ \hline
  & 0.1215(12) & \cite{BaSch} & 9.87(8) &\cite{Bali} & \\
 6.4  & 0.1218(28) & \cite{UKQCD_HW} &9.70(24) &\cite{UKQCD_HW} & \\
 & 0.1215(11) & Combined  &9.85(8) & Combined  &1.197(15) \\ \hline
 6.3 & 0.1394(11) & Interpolated & & & \\ \hline
 & 0.1610(9)  & \cite{HH_pers} &7.36(4) &\cite{Bali} & \\
  & 0.1608(23) & \cite{UKQCD_HW} &7.33(25) &\cite{UKQCD_HW} & \\[-0.6em]
6.2 & & & & & \\[-0.6em]
  & 0.1609(28) & \cite{CM_62} & & &  \\
 & 0.1610(8)  & Combined &7.36(4) &Combined & 1.185(9) \\ \hline
 6.17 & 0.1677(8) & Interpolated & & & \\ \hline
 & 0.2209(23) & \cite{HH_pers} &5.28(4) &\cite{Bali} &  \\
     & 0.2154(50) & \cite{UKQCD_HW} &5.53(15) &\cite{UKQCD_HW} & \\[-0.6em]
6.0 & & & & & \\[-0.6em]
     & 0.2182(21) & \cite{CM_60} & & & \\
 & 0.2191(15) & Combined &5.30(4) &Combined & 1.161(12)\\ \hline
 5.93 & 0.2536(29) & Interpolated & & & \\ \hline
 5.90 & 0.2702(37) & \cite{MTcK} &4.62(11) & \cite{Bali} & 1.25(3)\\ \hline
 5.85 & 0.2986(27) & Interpolated & & & \\ \hline
 5.8 & 0.3302(30) & \cite{MTcK} &3.63(5) &\cite{Bali} & 1.199(20)\\ \hline 
 5.7 & 0.4099(24) & \cite{MTcK} &2.86(5) &\cite{Bali} & 1.172(22) \\ \hline
 5.6 &  &  &2.29(6) &\cite{Bali} & \\ \hline
 5.5 &  &  &2.01(3) &\cite{Bali} & \\ \hline 
\end{tabular}
\end{center}
\vspace*{0.3cm}
\caption{The lattice spacing expressed in terms of the string tension 
$K$ and the force parameter $r_0$. When several groups have computed these 
quantities, we have taken the weighted average, while we interpolate
logarithmically whenever the values are not known.} 
\label{stringdata}
\end{table} 

\clearpage
\begin{table}
\begin{center}
\begin{tabular}{|l|l|} \hline
\multicolumn{1}{|c|}{$\beta$} &  \multicolumn{1}{c|}{$\kappa_c$}  \\ \hline
  6.40 &   0.150759(145) \\
  6.30 &   0.151774(36) \\
  6.20 &   0.153374(17) \\
  6.17 &   0.153838(37) \\
  6.00 &   0.157211(8) \\
  5.93 &   0.158985(73) \\
  5.85 &   0.161716(23) \\
  5.70 &   0.169313(72) \\ \hline
\end{tabular}
\end{center}
\vspace*{0.3cm}
\caption{The critical values of $\kappa$, $\kappa_c$, for the 
Wilson world data.}
\label{tkappa}
\end{table}

\begin{table}[hb]
\begin{center}
\begin{tabular}{|c|l|l|} \hline
\multicolumn{3}{|c|}{$\beta = 6.0$} \\ \hline
\multicolumn{3}{c}{ } \\[-0.8em] \hline
\multicolumn{3}{|c|}{$c_{SW} = 0$} \\ \hline 
$V$ & \multicolumn{1}{c|}{$\kappa$} & \multicolumn{1}{c|}{$2a\tilde{m}$} \\
\hline
    & 0.1487 & 0.2959(5) \\ 
    & 0.1515 & 0.1866(5) \\[-0.6em]  
$16^3 32$ & & \\[-0.6em]
    & 0.1530 & 0.1321(5) \\
    & 0.1550 & 0.0642(7) \\ \hline
\multicolumn{3}{c}{ }\\[-0.8em] \hline
\multicolumn{3}{|c|}{$c_{SW} = 1.769$}\\ \hline
$V$ & \multicolumn{1}{c|}{$\kappa$} & \multicolumn{1}{c|}{$2a\tilde{m}$} \\
\hline 
    & 0.1300 & 0.2836(3) \\ 
    & 0.1310 & 0.2279(3) \\
$16^3 32$ & 0.1324 & 0.15231(10) \\
    & 0.1333 & 0.10380(11) \\
    & 0.1342 & 0.0553(2) \\ \hline 
    & 0.1342 & 0.0551(3) \\
$24^3 32$ & 0.1346 & 0.0330(3) \\
    & 0.1348 & 0.0214(4) \\ \hline 
\end{tabular}
\end{center}
\vspace*{0.3cm}
\caption{The bare quark masses $\tilde{m}$ for 
Wilson fermions ($c_{SW}=0$) and improved fermions 
($c_{SW}=1.769$) at $\beta = 6.0$. The errors are bootstrap errors.}
\label{tward1}
\end{table}

\clearpage
\begin{table}
\begin{center}
\begin{tabular}{|c|l|l|} \hline
\multicolumn{3}{|c|}{$\beta = 6.2$} \\ \hline
\multicolumn{3}{c}{ } \\[-0.8em] \hline
\multicolumn{3}{|c|}{$c_{SW} = 0$} \\ \hline 
$V$ & \multicolumn{1}{c|}{$\kappa$} & \multicolumn{1}{c|}{$2a\tilde{m}$} \\
\hline
    & 0.1468 & 0.2474(3) \\ 
    & 0.1489 & 0.1616(3) \\
$24^3 48$ & 0.1509 & 0.0845(3) \\
    & 0.1518 & 0.0514(3) \\
    & 0.1523 & 0.0336(3) \\ \hline 
\multicolumn{3}{c}{ }\\[-0.8em] \hline
\multicolumn{3}{|c|}{$c_{SW} = 1.614$}\\ \hline
$V$ & \multicolumn{1}{c|}{$\kappa$} & \multicolumn{1}{c|}{$2a\tilde{m}$} \\
\hline 
    & 0.1321 & 0.2161887) \\ 
    & 0.1333 & 0.14585(7) \\
$24^3 48$ & 0.1344 & 0.08185(7) \\
    & 0.1349 & 0.05283(8) \\
    & 0.1352 & 0.03538(9) \\ \hline 
\end{tabular}
\end{center}
\vspace*{0.3cm}
\caption{The bare quark masses $\tilde{m}$ for 
Wilson fermions ($c_{SW}=0$) and improved fermions 
($c_{SW}=1.614$) at $\beta = 6.0$. The errors are bootstrap errors.}
\label{tward2}
\end{table}

\begin{table}[hb]
\begin{center}
\begin{tabular}{|c|c|r|r|r|r|} \hline
   & & \multicolumn{2}{c|}{ } & 
\multicolumn{2}{c|}{ } \\[-0.8em] 
   & & \multicolumn{2}{c|}{$m_{u,d}^{\overline{MS}}$} & 
\multicolumn{2}{c|}{$m_s^{\overline{MS}}$} \\[-0.3em] 
\multicolumn{1}{|c|}{$\beta$} &\multicolumn{1}{|c|}{$c_{SW}$} & 
\multicolumn{2}{c|}{ } &
\multicolumn{2}{c|}{ } \\[-0.6em]
   & & \multicolumn{1}{c|}{Ward} & \multicolumn{1}{c|}{Standard} & 
\multicolumn{1}{c|}{Ward} &  \multicolumn{1}{c|}{Standard} \\  \hline    
6.0 & 0   & $4.40 \pm 0.17$ & $6.47 \pm 0.20$ 
& $105.0 \pm 4.5$ & $141.8 \pm 6.0$ \\ 
6.2 & 0   & $4.73 \pm 0.14$ & $6.39 \pm 0.25$ 
& $108.5 \pm 4.2$ & $138.8 \pm 7.4$
\\ \hline
6.0 & 1.769   & $4.02 \pm 0.10$ & $4.94 \pm 0.93$ & $92.8 \pm 2.9$ & $109.4 \pm 3.9$
\\ 
6.2 & 1.614   & $4.47 \pm 0.06$ & $5.09 \pm 0.16$ & $101.6 \pm 1.7$ & $111.7 \pm 4.7$
\\ 
$\infty$  & & $5.00 \pm 0.18$ & $5.27 \pm 0.36$ & $111.9 \pm 5.0$ & 
$114.4 \pm 11.1$ \\ \hline
\end{tabular}
\end{center}
\vspace*{0.3cm}
\caption{Our results of the renormalized quark masses 
$m^{\overline{MS}}(2\, \mbox{GeV})$ 
in $\mbox{MeV}$ for improved and Wilson fermions, together with the 
extrapolation to the continuum limit ($\beta = \infty$). The continuum 
numbers refer to improved fermions. We give the results for both the Ward
identity and the standard method.}
\label{mMeV}
\end{table}

\clearpage
\begin{table}
\begin{center}
\begin{tabular}{|c|l|l|l|l|l|l|}
\hline 
\multicolumn{7}{|c|}{$\beta = 6.0$} \\
\hline
\multicolumn{7}{c}{ } \\[-0.8em]
\hline
\multicolumn{7}{|c|}{$c_{SW} = 0$} \\
\hline
$V$ &
   \multicolumn{1}{c|}{$\kappa$} &
   \multicolumn{1}{c|}{$af^{(0)}_\pi$} &
   \multicolumn{1}{c|}{$ - $} &
   \multicolumn{1}{c|}{$f^{(0)}_\rho$} &
   \multicolumn{1}{c|}{$ - $} &
   \multicolumn{1}{c|}{$f^{(0)}_{a_1}$}  \\
\hline
    & 0.1487 & 0.136(2)  &  & 0.305(5) &  & 0.161(5) \\ 
    & 0.1515 & 0.122(2)  &  & 0.364(5) &  & 0.207(3) \\[-0.6em]
$16^3 32$ & & & & & & \\[-0.6em]
    & 0.1530 & 0.113(2)  & & 0.397(7) &  & 0.231(3) \\
    & 0.1550 & 0.098(2)  & & 0.459(9) &  & 0.262(4) \\
\hline
\multicolumn{7}{c}{ } \\[-0.8em]
\hline
\multicolumn{7}{|c|}{$c_{SW} = 1.769$} \\
\hline
$V$ &
   \multicolumn{1}{c|}{$\kappa$} &
   \multicolumn{1}{c|}{$af^{(0)}_\pi$} &  
   \multicolumn{1}{c|}{$a f^{(1)}_\pi/f^{(0)}_\pi$} &
   \multicolumn{1}{c|}{$f^{(0)}_\rho$} &
   \multicolumn{1}{c|}{$a f^{(1)}_\rho/f^{(0)}_\rho$} &
   \multicolumn{1}{c|}{$f^{(0)}_{a_1}$}  \\
\hline
    & 0.1300 & 0.1341(15)& 1.792(7) & 0.209(9) & 0.670(2)  & 0.131(7) \\
    & 0.1310 & 0.1295(15)& 1.698(8) & 0.228(3) & 0.588(2)  & 0.153(13)\\
$16^3 32$
    & 0.1324 & 0.1204(8) & 1.599(5) & 0.261(2) & 0.4823(12)& 0.172(7) \\
    & 0.1333 & 0.1128(9) & 1.541(3) & 0.288(3) & 0.4147(15)& 0.202(16)\\ 
    & 0.1342 & 0.1037(8) & 1.511(6) & 0.323(3) & 0.353(2)  & 0.208(16)\\
\hline
    & 0.1342 & 0.105(2)  & 1.521(16)& 0.330(7) & 0.348(3)  & 0.212(10)\\
$24^3 32$
    & 0.1346 & 0.101(2)  & 1.58(3)  & 0.352(7) & 0.327(6)  & 0.225(14)\\
    & 0.1348 & 0.100(3)  & 1.62(5)  & 0.352(15)& 0.324(19) & 0.25(2)  \\
\hline
\end{tabular}
\end{center}
\vspace*{0.3cm}
\caption{The various contributions to the decay constants $f_\pi$, $f_\rho$
and $f_{a_1}$ at $\beta=6.0$.}
\label{decay60}
\end{table}

\clearpage
\begin{table}
\begin{center}
\begin{tabular}{|c|l|l|l|l|l|l|}
\hline 
\multicolumn{7}{|c|}{$\beta = 6.2$} \\
\hline
\multicolumn{7}{c}{ } \\[-0.8em]
\hline
\multicolumn{7}{|c|}{$c_{SW} = 0$} \\
\hline
$V$ &
   \multicolumn{1}{c|}{$\kappa$} &
   \multicolumn{1}{c|}{$af^{(0)}_\pi$} &  
   \multicolumn{1}{c|}{$ - $} &
   \multicolumn{1}{c|}{$f^{(0)}_\rho$} &
   \multicolumn{1}{c|}{$ - $} &
   \multicolumn{1}{c|}{$f^{(0)}_{a_1}$}  \\
\hline
    & 0.1468 & 0.1025(19)&  & 0.268(5) &  & 0.127(9) \\
    & 0.1489 & 0.0930(17)&  & 0.315(6) &  & 0.180(4) \\
$24^3 48$
    & 0.1509 & 0.0798(14)&  & 0.376(7) &  & 0.230(4) \\
    & 0.1518 & 0.0719(13)&  & 0.412(9) &  & 0.261(4) \\
    & 0.1523 & 0.0669(14)&  & 0.438(10)&  & 0.276(5) \\
\hline
\multicolumn{7}{c}{ } \\[-0.8em]
\hline
\multicolumn{7}{|c|}{$c_{SW} = 1.614$} \\
\hline
$V$ &
   \multicolumn{1}{c|}{$\kappa$} &
   \multicolumn{1}{c|}{$af^{(0)}_\pi$} &  
   \multicolumn{1}{c|}{$a f^{(1)}_\pi/f^{(0)}_\pi$} &
   \multicolumn{1}{c|}{$f^{(0)}_\rho$} &
   \multicolumn{1}{c|}{$a f^{(1)}_\rho/f^{(0)}_\rho$} &
   \multicolumn{1}{c|}{$f^{(0)}_{a_1}$}  \\
\hline
    & 0.1321 & 0.0985(11)& 1.297(3) & 0.211(3) & 0.4637(8) & 0.133(2) \\
    & 0.1333 & 0.0913(11)& 1.198(4) & 0.243(3) & 0.3719(10)& 0.167(2) \\
$24^3 48$
    & 0.1344 & 0.0818(10)& 1.133(2) & 0.283(4) & 0.2900(15)& 0.204(3) \\
    & 0.1349 & 0.0758(9) & 1.118(7) & 0.308(5) & 0.255(2)  & 0.226(3) \\
    & 0.1352 & 0.072(3)  & 1.131(11)& 0.327(16)& 0.235(4)  & 0.241(11)\\
\hline
\end{tabular}
\end{center}
\vspace*{0.3cm}
\caption{The same as table \ref{decay60} but for $\beta=6.2$.}
\label{decay62}
\end{table}

\begin{table}
\begin{center}
\begin{tabular}{|c|l|l|l|l|l|l|}
\hline 
$\beta$ &
   \multicolumn{1}{c|}{$c_{SW}$} &
   \multicolumn{1}{c|}{$af_\pi$} &  
   \multicolumn{1}{c|}{$f_{a_1}$} &
   \multicolumn{1}{c|}{$1/f_\rho$} &
   \multicolumn{1}{c|}{$af_K$} &  
   \multicolumn{1}{c|}{$1/f_{K^*}$}  \\
\hline
$6.0$ & $0$   & 
     0.0569(77) & 0.2240(76)  & 0.295(11) & 0.0732(26)  & 0.2385(24) \\
$6.2$ & $0$   & 
     0.0423(36) & 0.2429(48) & 0.2971(71) & 0.0537(11)  & 0.2345(28) \\ \hline
$6.0$ & $1.769$ &
     0.0627(20) & 0.195(10) & 0.2664(39)  & 0.0721(8) & 0.2020(13) \\ 
$6.2$ & $1.614$ & 
     0.0462(38) & 0.2130(45) & 0.2726(74) & 0.0557(14)  & 0.2149(19) \\
\hline
\end{tabular}
\end{center}
\vspace*{0.3cm}
\caption{The decay constants $f_\pi$, $f_{a_1}$, $f_\rho$ extrapolated
to the chiral limit, as well as $f_K$, $f_{K^*}$ taken at the physical quark
mass.}
\label{decaychi}
\end{table}

\clearpage
\section*{Figures}

\clearpage

\clearpage

\clearpage

\clearpage

\clearpage

\clearpage

\clearpage

\clearpage

\clearpage

\clearpage

\clearpage

\clearpage

\clearpage

\clearpage
\begin{figure}
\begin{centering}
\epsfig{figure=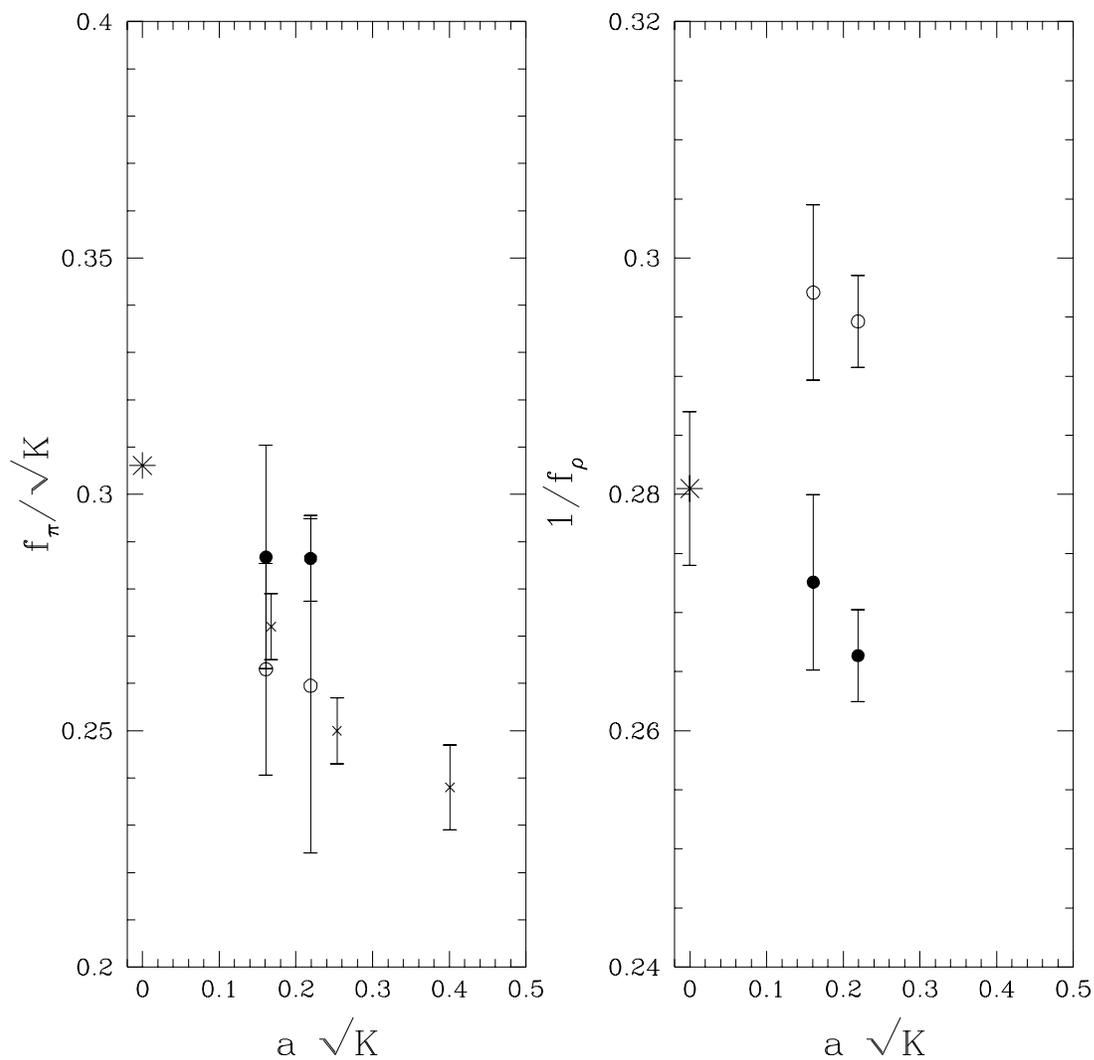,height=15.0cm,width=15.0cm}
\caption{The decay constants $f_\pi$ and $f_\rho$ as a function of
the lattice spacing for improved ($\kreiso$) and Wilson ($\kreisv$) fermions
together with the experimental values ($\stern$). The errors on 
$f_\rho$ for improved fermions are statistical only.
Our results for $f_\pi$ are compared with the Wilson results of 
ref.~\protect\cite{wein} ($\times$).}
\end{centering}
\label{figpirho}
\end{figure}

\clearpage
\begin{figure}
\begin{centering}
\epsfig{figure=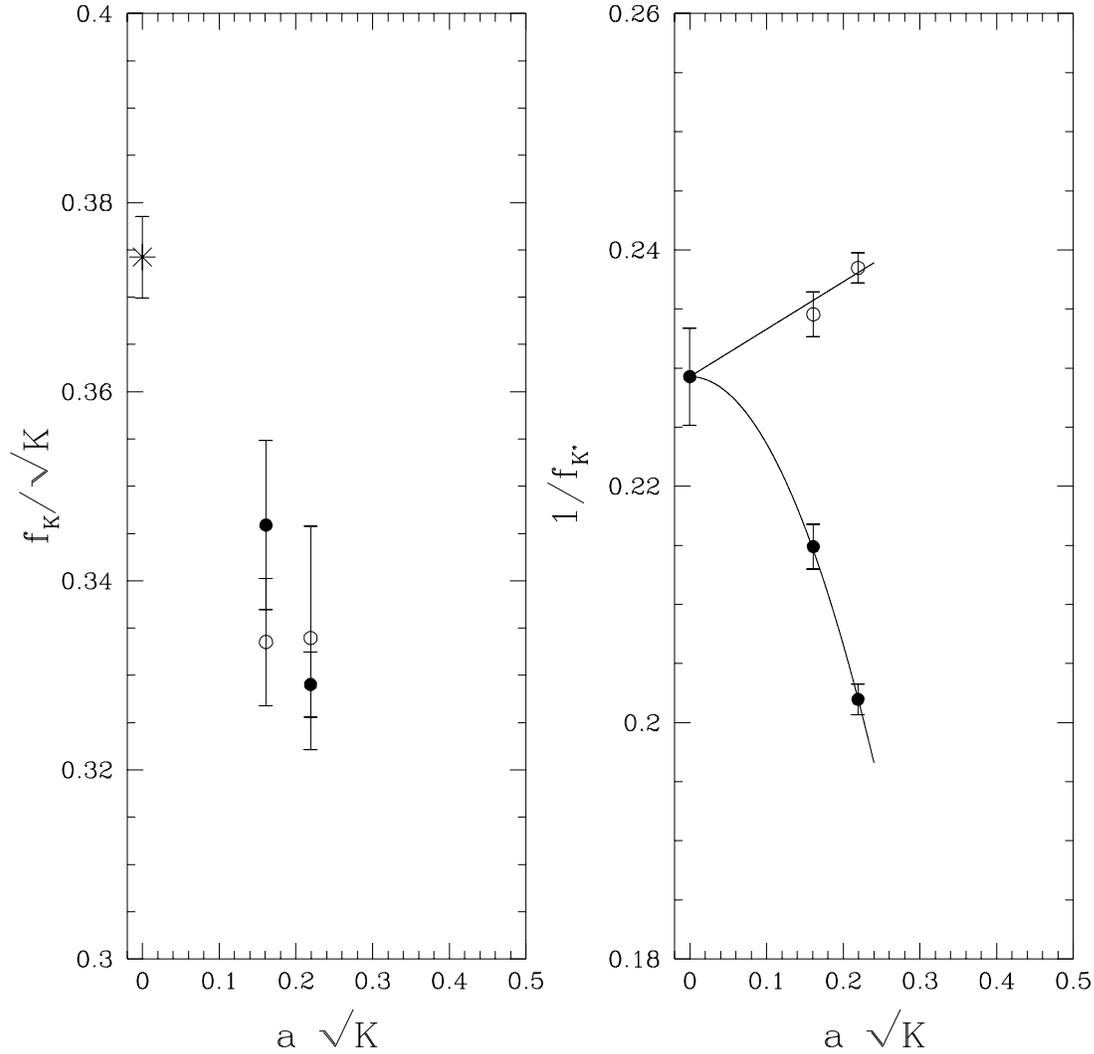,height=15.0cm,width=15.0cm}
\caption{The decay constants $f_K$ and $f_{K^*}$ as a function of the
lattice spacing for improved ($\kreiso$) and Wilson ($\kreisv$)
fermions. The errors on $f_{K^*}$ for improved fermions are statistical only.
The solid lines in the $f_{K^*}$ figure are from a simultaneous 
linear fit to the Wilson data and a quadratic fit to the improved data. Our
results for $f_K$ are compared with the experimental value ($\stern$).}
\end{centering}
\label{figkkstern}
\end{figure}

\end{document}